\documentclass[twocolumn, preprint]{aastex62}
\usepackage{hyperref}
\usepackage{multirow}

\begin{document}

\title{CO~($J=1-0$) Observations toward Filamentary Molecular Clouds in the Galactic Region with $l = [169\arcdeg.75, 174\arcdeg.75], b = [-0\arcdeg.75, 0\arcdeg.5]$}

\author{Fang Xiong}
\affil{Purple Mountain Observatory, and Key Laboratory of Radio Astronomy, Chinese Academy of Sciences, 8 Yuanhua Road, Nanjing 210034, China; fangxiong@pmo.ac.cn, xpchen@pmo.ac.cn}
\affil{University of Chinese Academy of Sciences, 19A Yuquan Road, Shijingshan District, Beijing 100049, China}
\affil{Harvard-Smithsonian Center for Astrophysics, 60 Garden Street, Cambridge, MA 02138, USA}

\author{Xuepeng Chen}
\affil{Purple Mountain Observatory, and Key Laboratory of Radio Astronomy, Chinese Academy of Sciences, 8 Yuanhua Road, Nanjing 210034, China; fangxiong@pmo.ac.cn, xpchen@pmo.ac.cn}
\affil{School of Astronomy and Space Science, University of Science and Technology of China, Hefei, Anhui 230026, China}

\author{Qizhou Zhang}
\affil{Harvard-Smithsonian Center for Astrophysics, 60 Garden Street, Cambridge, MA 02138, USA}

\author{Ji Yang}
\affil{Purple Mountain Observatory, and Key Laboratory of Radio Astronomy, Chinese Academy of Sciences, 8 Yuanhua Road, Nanjing 210034, China; fangxiong@pmo.ac.cn, xpchen@pmo.ac.cn}
\affil{School of Astronomy and Space Science, University of Science and Technology of China, Hefei, Anhui 230026, China}

\author{Min Fang}
\affil{Department of Astronomy, University of Arizona, 933 North Cherry Avenue, Tucson, AZ 85721, USA}

\author{Miaomiao Zhang}
\affil{Purple Mountain Observatory, and Key Laboratory of Radio Astronomy, Chinese Academy of Sciences, 8 Yuanhua Road, Nanjing 210034, China; fangxiong@pmo.ac.cn, xpchen@pmo.ac.cn}
\affil{Max-Planck-Institut f{\"u}r Astronomie, K{\"o}nigstuhl 17, D-69117 Heidelberg, Germany}

\author{Weihua Guo}
\author{Li Sun}
\affil{Purple Mountain Observatory, and Key Laboratory of Radio Astronomy, Chinese Academy of Sciences, 8 Yuanhua Road, Nanjing 210034, China; fangxiong@pmo.ac.cn, xpchen@pmo.ac.cn}
\affil{School of Astronomy and Space Science, University of Science and Technology of China, Hefei, Anhui 230026, China}

\begin{abstract}
We present observations of the CO isotopologues ($^{12}$CO, $^{13}$CO, and C$^{18}$O) toward the Galactic region with $169\arcdeg.75 \leqslant l \leqslant 174\arcdeg.75$ and $-0\arcdeg.75 \leqslant b \leqslant 0\arcdeg.5$, using the Purple Mountain Observatory 13.7~m millimeter-wavelength telescope. Based on the $^{13}$CO~($J = 1-0$) data, we find five molecular clouds within the velocity range between $-$25 and 8~km~s$^{-1}$ that are all characterized by conspicuous filamentary structures. We have identified eight filaments with a length of 6.38--28.45~pc, a mean H$_2$ column density of 0.70$\times$10$^{21}$--6.53$\times$10$^{21}$~cm$^{-2}$, and a line mass of 20.24--161.91~$M_\sun$ pc$^{-1}$, assuming a distance of $\sim$1.7~kpc. Gaussian fittings to the inner parts of the radial density profiles lead to a mean FWHM width of 1.13$\pm$0.01~pc. The velocity structures of most filaments present continuous distributions with slight velocity gradients. We find that turbulence is the dominant internal pressure to support the fragmentation of filaments instead of thermal pressure. Most filaments have virial parameters smaller than 2; thus, they are gravitationally bound. Four filaments have an LTE line mass close to the virial line mass. We further extract dense clumps using the $^{13}$CO data and find that 64$\%$ of the clumps are associated with the filaments. According to the complementary IR data, most filaments have associated Class~II young stellar objects. Class~I objects are mainly found to be located in the filaments with a virial parameter close to 1. Within two virialized filaments, $^{12}$CO outflows have been detected, indicating ongoing star-forming activity therein.
\end{abstract}


\section{INTRODUCTION}
Filamentary structures are ubiquitous in molecular clouds and have been known for several decades \citep{sch79, bal87, lor89, abe94, hat05, mye09}. The mid-IR observations conducted by the $Midcourse$ $Space$ $Experiment$ \citep[$MSX$;][]{car98, car00} and $Spitzer$ \citep{chu09, per09} and submillimeter observations conducted by $Herschel$ \citep{and10, men10, mol10} and $Planck$ \citep{pla15, pla16} have also demonstrated the omnipresence of filamentary structures in the infrared dark clouds (IRDCs) and cold interstellar medium (ISM). Numerous studies employed the data from previous surveys to identify the large-scale filamentary structures within the Milky Way. \citet{rag14} made use of Galactic Ring Survey \citep[GRS;][]{jac06} data and identified seven giant filaments with a length of $\sim$100~pc in the first Galactic quadrant. \citet{wan15} found nine prominent filaments with a length of 37--99~pc in the region with $15\arcdeg < l < 56\arcdeg, |b| \leqslant 1\arcdeg$ using the $Herschel$ Infrared Galactic Plane Survey \citep[Hi-GAL;][]{mol10}. \citet{zuc15} investigated the region ($|l| < 62\arcdeg, |b| < 1\arcdeg$) covered by the $Spitzer$ MIPS Galactic Plane Survey \citep[MIPSGAL;][]{car09} and found 10 ``bone'' filaments with a length of 13--52~pc. \citet{lig16} identified 486 spatially and kinematically coherent filaments with a length of 2--20~pc in the inner Galactic plane based on the APEX Telescope Large Area Survey of the Galaxy \citep[ATLASGAL;][]{sch09}.

As pointed out by \citet{and10} and \citet{men10}, there is an intimate correspondence between the spatial distribution of dense cores and the filamentary network. Theoretical works have shown that filamentary molecular clouds can fragment into clumps as a result of gravitational instability \citep{nak93, tom95, van14}, and some of these clumps have sufficient density to evolve, collapse, and form stars \citep{con16}. \citet{hen10} found that about 75$\%$ of the pre- and protostellar cores detected by $Herschel$ PACS observations are located within the filamentary IRDC G011.11$-$0.12. \citet{kon15} also found that about 78$\%$ of prestellar cores lie within a 0.1~pc width \citep{arz11} of the filaments' footprints traced by the DisPerSE algorithm \citep{sou11} in the Aquila cloud. It is reasonable to believe that filamentary structures, as the most common shape of molecular clouds, may be the preferred birthplace of dense clumps and prestellar cores \citep{and10, bus13, lux18}.

Magnetic fields also play an important role in the formation of filamentary structures. \citet{ino13} and \citet{ino18} proposed a scenario that a filament is formed due to the convergence of an inhomogeneous cloud, which is caused by the bending of the ambient magnetic field, when sweeping by a magnetohydrodynamics (MHD) shock wave. This scenario is compatible with the CO observations toward the filamentary molecular cloud in Taurus \citep{arz18}. In addition, magnetic fields also help to shape and stabilize the filaments \citep{pal13, kir15}. \citet{kus16} found that the magnetic field appears to be perpendicular to the inner ``ridge filaments'' in Vela C. The similar magnetic field running perpendicular to the filament is also found in the B211 region of Taurus \citep{pal13} and IRDC G14.225$-$0.506 \citep{bus13, san16}. \citet{arz19} pointed out that thermally supercritical filaments tend to be mostly perpendicular to the magnetic field lines and are the main sites of prestellar core formation \citep{and10, kon15}, while thermally subcritical filaments are well aligned with the local magnetic field lines and without any star forming activity \citep{arz13, hac16}.

Hence, it is of great importance to study various filamentary structures (e.g., small-scale and large-scale, low-mass and high-mass) within the Milky Way, characterizing their properties in detail and understanding the star-forming activity therein. We have conducted molecular line observations toward two Galactic regions centered at $l = 150\arcdeg, b = 3\arcdeg.5$ (referred to hereafter as the ``G150 region'') and $l = 172\arcdeg, b = -0\arcdeg.2$ (referred to hereafter as the ``G172 region''), both of which have significant filamentary structures within. Our work is part of the Milky Way Imaging Scroll Painting (MWISP) project,\footnote{\url{http://english.dlh.pmo.cas.cn/ic/in/} and \url{http://www.radioast.nsdc.cn/mwisp.php}} conducted by the Purple Mountain Observatory (PMO) using the PMO 13.7~m radio telescope to investigate the molecular gas along the northern Galactic plane with the $^{12}$CO, $^{13}$CO, and C$^{18}$O~($J = 1-0$) lines. Our work in the G150 region \citep[see][]{xio17} showed that the molecular gas in this region consists of one main filament in the inner area with a higher column density and 11 subfilaments in the outer area with lower column densities, forming the so-called ``ridge-nest'' structure. About 75$\%$ of the CO dense clumps are associated with the filaments, and 56$\%$ of the virialized clumps are associated with the gravitationally bound filaments. For the G172 region, our observations reveal that most of the molecular clouds in this region present filamentary structures (see Figure~\ref{fig:f1}). However, so far, there has been no detailed study of these filamentary molecular clouds, and the properties of the filaments and dense clumps within are still unknown. 

In this paper, we present the MWISP CO~($J = 1-0$) observations toward the G172 region in the CO isotopologues. With the multiline observations (the angular resolution of the MWISP survey is $\sim$50$\arcsec$ and the velocity resolution is $\sim$0.16~km~s$^{-1}$), we identify five different velocity components characterized by filamentary structures, study in detail their dynamical properties, and investigate the molecular clumps and young stellar objects (YSOs) associated with them. In Section~\ref{sec:obs}, we describe the CO line observations and data reduction. We present our main results in Section~\ref{sec:res} and discussions in Section~\ref{sec:disc}. The conclusions are summarized in Section~\ref{sec:sum}.

\section{OBSERVATIONS AND DATA REDUCTION}\label{sec:obs}

\subsection{CO Data}
Our observations of the G172 region covered an area within $169\arcdeg.75 \leqslant l \leqslant 174\arcdeg.75$ and $-0\arcdeg.75 \leqslant b \leqslant 0\arcdeg.5$. The operation of observations was conducted by the 13.7~m millimeter-wavelength PMO telescope located in Delingha, China, from 2015 November to 2016 December. The $^{12}$CO~($J = 1-0$), $^{13}$CO~($J = 1-0$), and C$^{18}$O~($J = 1-0$) emission lines were observed simultaneously with the nine-beam Superconducting Spectroscopic Array Receiver \citep[SSAR;][]{sha12}. The half-power beamwidth of the telescope was about 52$\arcsec$ at 115.2~GHz, and 55$\arcsec$ at 110.2~GHz.

The telescope scanned the sky along both the longitude and latitude directions at a constant rate of 50$\arcsec$~s$^{-1}$, using the on-the-fly (OTF) observation mode. During the observations, the typical system temperature was 270~K for the $^{12}$CO line and 175~K for the $^{13}$CO and C$^{18}$O lines. The antenna temperature ($T_{\rm A}^*$) was calibrated to the main beam temperature ($T_{\rm MB}$) by using $T_{\rm MB} = T_{\rm A}^*/\eta_{\rm MB}$, with a main beam efficiency ($\eta_{\rm MB}$) of 44$\%$ for the $^{12}$CO line and 48$\%$ for the $^{13}$CO and C$^{18}$O lines. 

The bad channels and standing waves were removed when checking the spectra after we derived the original OTF data. Using the GILDAS software package \citep{gui00, pet05, gil13}, we regridded the data into $30\arcsec \times 30\arcsec$ pixels (which is approximately half of the beam size based on the unified data reduction for the MWISP project) and converted them into the standard FITS files. We further used the IDL software to mosaic these FITS files together to generate the final data cube of the G172 region. As a result, the rms noise level is 0.49~K for the $^{12}$CO and 0.26~K for the $^{13}$CO and C$^{18}$O, and the velocity resolution is 0.16~km~s$^{-1}$ for the $^{12}$CO and 0.17~km~s$^{-1}$ for the $^{13}$CO and C$^{18}$O.

For the $^{12}$CO and C$^{18}$O emission lines, we further applied Gaussian fitting to the spectra using the IDL software. The $^{13}$CO~($J = 1-0$) line has two hyperfine lines, which are $^{13}$CO~($J = 1-0, F = 1/2-1/2$) ($\nu$=110.20132180~GHz) and $^{13}$CO~($J = 1-0, F = 3/2-1/2$) ($\nu$=110.20137040~GHz). We used the hyperfine line model from the Python package to fit the hyperfine structures of $^{13}$CO. The fitting results are presented in Section~\ref{sec:res}.

\subsection{IR Data}
In this work, we used the complementary IR data from the Two Micron All Sky Survey \citep[2MASS;][]{skr06} with near-IR $J$ (1.25~$\mu$m), $H$ (1.65~$\mu$m), and $K_{\rm s}$ (2.16~$\mu$m) bands; the $Wide-field$ $Infrared$ $Survey$ $Explorer$ \citep[$WISE$;][]{wri10} with four IR bands,  $W1$ (3.4~$\mu$m), $W2$ (4.6~$\mu$m), $W3$ (12~$\mu$m), and $W4$ (22~$\mu$m); and the $Spitzer$ \citep{wer04} Galactic Legacy Infrared Mid-Plane Survey Extraordinaire \citep[GLIMPSE;][]{ben03, chu09, whi11} with two IRAC bands, [3.6~$\mu$m] and [4.5~$\mu$m]. The IR data were retrieved from the NASA/IPAC Infrared Science Archive (IRSA)\footnote{\url{http://irsa.ipac.caltech.edu/frontpage/}}.

\section{RESULTS}\label{sec:res}

\subsection{Overview and Identification of Filamentary Structures}
\begin{figure}
\epsscale{1.0}
\plotone{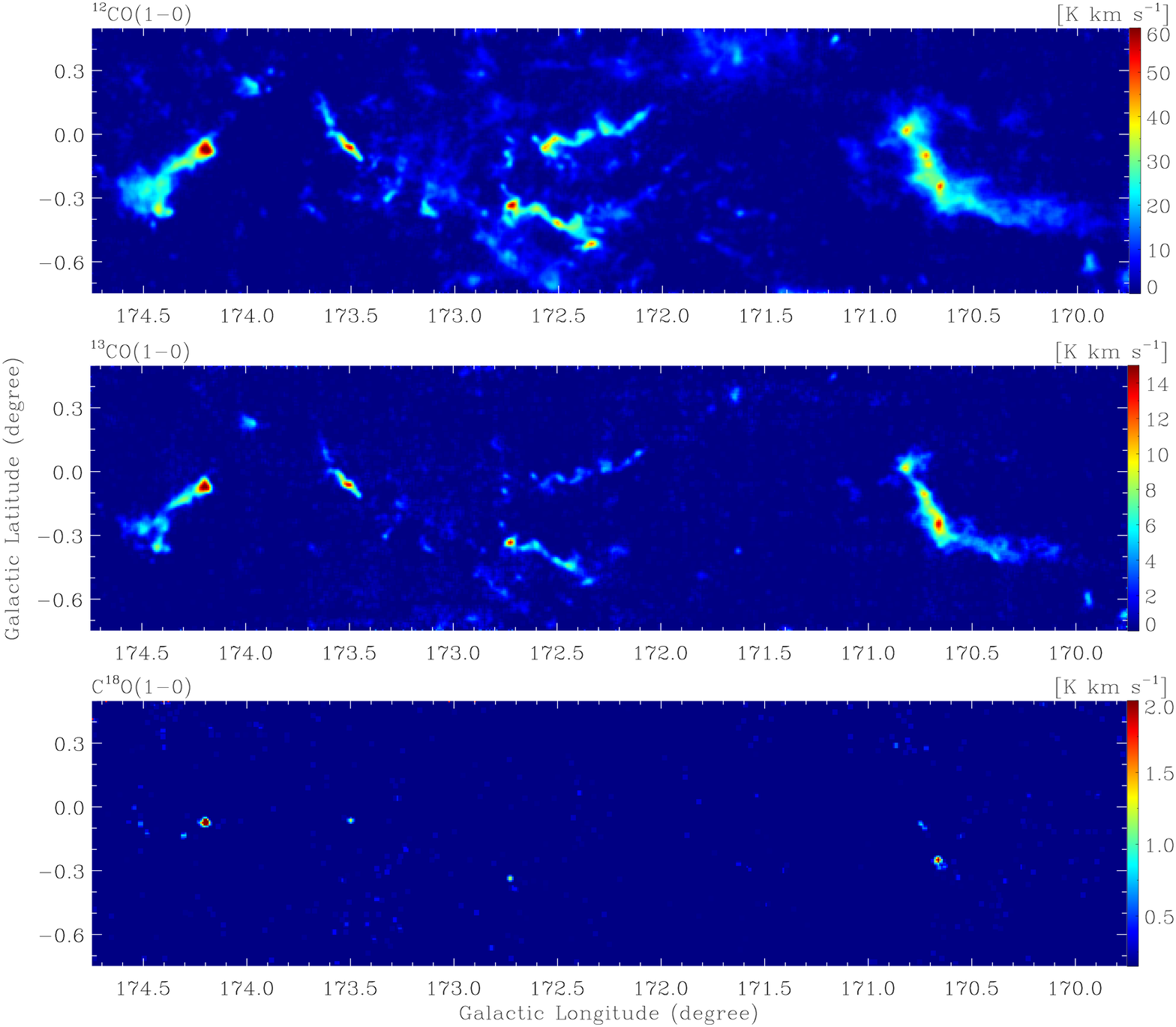}
\caption{Top panel: integrated intensity map of the G172 region in the $^{12}$CO emission with a velocity range from $-$24.5 to 7.5~km~s$^{-1}$. Middle panel: integrated intensity map of the G172 region in the $^{13}$CO emission with a velocity range from $-$24.5 to 7.5~km~s$^{-1}$. Bottom panel: integrated intensity map of the G172 region in the C$^{18}$O emission with a velocity range from $-$24.5 to 7.5~km~s$^{-1}$.}
\label{fig:f1}
\end{figure}

As shown in Figure~\ref{fig:f1}, our CO observations of the G172 region reveal several filamentary molecular clouds in this region. The CO emission within this region covers the velocity range from $-$24.5 to 7.5~km~s$^{-1}$. Due to isotopic abundance ratios, the transitions of rare isotopologues (e.g., $^{12}$CO~($J = 1-0$), $^{13}$CO~($J = 1-0$), and C$^{18}$O~($J = 1-0$)) would have smaller optical depths (see Section 3.2) at the same densities. Moreover, the $J = 1-0$ transitions of $^{12}$CO, $^{13}$CO, and C$^{18}$O have slightly different excitation conditions \citep[e.g.,][]{man15}. The different excitation conditions, along with the different optical depths, would further result in the $J = 1-0$ transitions of various CO isotopologues being able to trace different gas densities. The $^{12}$CO emission traces the outlines of filamentary structures with a density of $\sim$10$^{2}$~cm$^{-3}$, while the $^{13}$CO emission reveals the skeletons of filamentary structures with a typical density of $\sim$10$^{3}$~cm$^{-3}$. The C$^{18}$O emission marks the densest parts of the $^{13}$CO emission with a higher density of $\sim$10$^{4}$~cm$^{-3}$. The two filamentary clouds in the outer area (the one in the east and the one in the west) have stronger integrated intensities than those in the inner area. The emission of the C$^{18}$O line is also mainly located in the outer area.

\begin{figure}
\epsscale{1.0}
\plotone{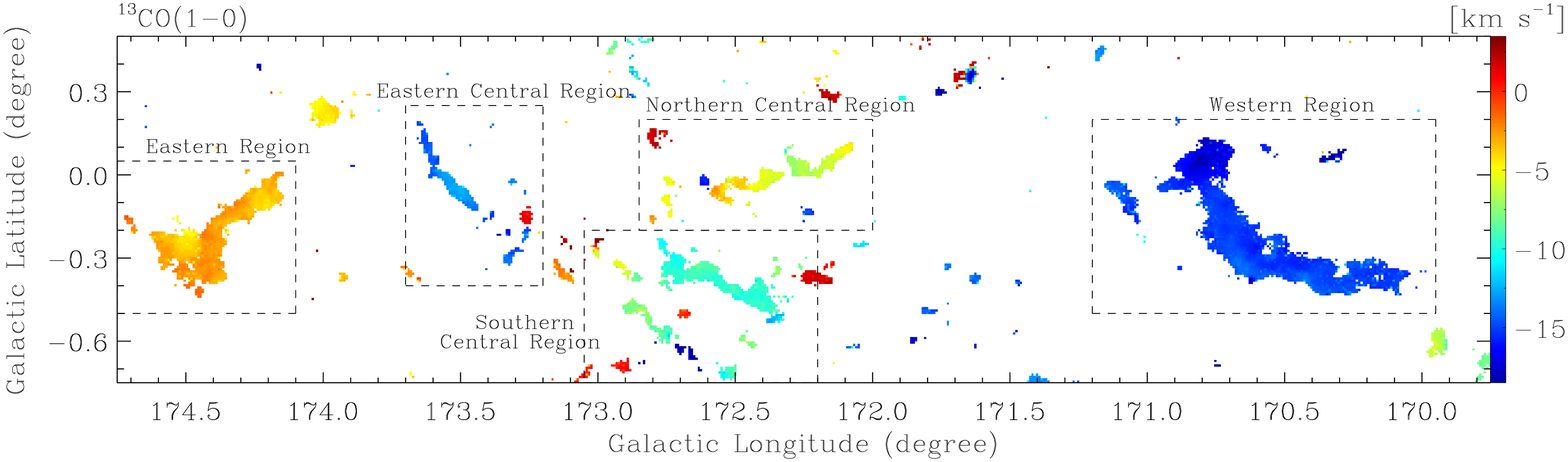}
\caption{Velocity distribution map of molecular clouds in the G172 region traced by the $^{13}$CO emission. The velocity range is from $-$18.5 to 3.5~km~s$^{-1}$. The name of each region is marked beside it.}
\label{fig:f2}
\end{figure}

\begin{figure}
\epsscale{1.0}
\plotone{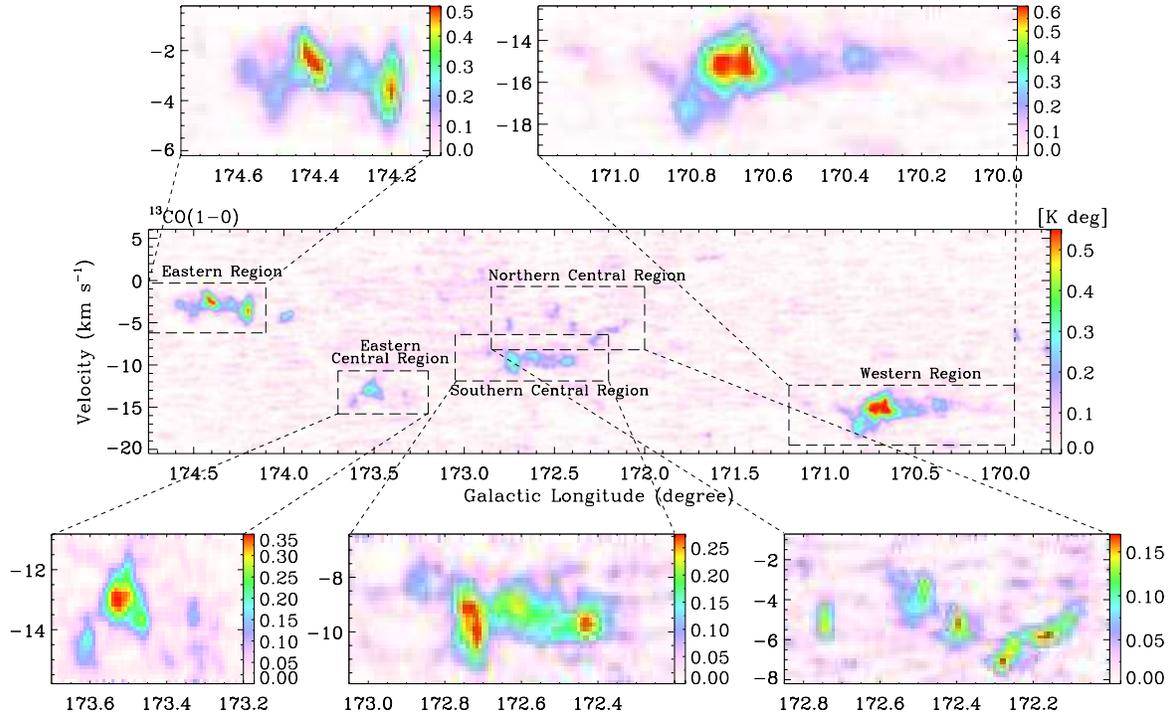}
\caption{Top panels: longitude-velocity maps of the eastern and western regions with the integrated latitude ranges of $-0\arcdeg.5$ to $0\arcdeg.05$ and $-0\arcdeg.5$ to $0\arcdeg.2$, respectively. Middle panel: longitude-velocity map of the whole G172 region with the integrated latitude range of $-0\arcdeg.75$ to $5\arcdeg$. Bottom panels: longitude-velocity maps of the eastern central, southern central, and northern central regions with the integrated latitude ranges of $-0\arcdeg.4$ to $0\arcdeg.25$, $-0\arcdeg.75$ to $-0\arcdeg.2$, and $-0\arcdeg.2$ to $0\arcdeg.2$, respectively.}
\label{fig:f3}
\end{figure}

Based on the $^{13}$CO data, we derived the velocity distribution map shown in Figure~\ref{fig:f2}. The filamentary molecular clouds in this region cover a wide range of velocity components, ranging from $-$15~km~s$^{-1}$ (the one in the west\footnote{The direction words in this work refer to the Galactic coordinate.}) to $-$2~km~s$^{-1}$ (the one in the east). To illustrate the velocity components more directly, we derived a longitude-velocity map of this region, shown in the middle panel of Figure~\ref{fig:f3}. It is evident that these filamentary molecular clouds have quite different velocity components. According to the difference in velocity and spatial distributions (see the middle panel of Figure~\ref{fig:f1}), we divide the filamentary molecular clouds of the G172 region into five main subregions--i.e., the eastern region, eastern central region, northern central region, southern central region, and western region--in order to study their properties separately and in detail. The velocity components of these five subregions are presented in the top and bottom panels of Figure~\ref{fig:f3}.

As defined by \citet{and14}, a ``filament'' is any elongated ISM structure with an aspect ratio larger than $\sim$5-10 that is significantly overdense with respect to its surroundings. They also pointed out that filaments are generally linear over their length and appear to be colinear in the direction of the longer extents of their host clouds. \citet{hac17, hac18} further confined the velocity coherence of the filaments by adopting the local measured line width as the velocity threshold to check the points along the filaments. According to these definitions, we adopt the ``FilFinder'' algorithm \citep{koc15} to identify filamentary structures within each subregion. FilFinder is a new approach to identifying filaments within molecular clouds based on the application of mathematical morphology. As illustrated in \citet{koc15}, FilFinder relies on an adaptive threshold, which will create a mask based on local changes in the brightness of the intensity image. The medial axis transform will then reduce the signal mask to a skeleton. This skeleton will further be pruned down to form a filamentary network.

\begin{figure}
\epsscale{1.0}
\plotone{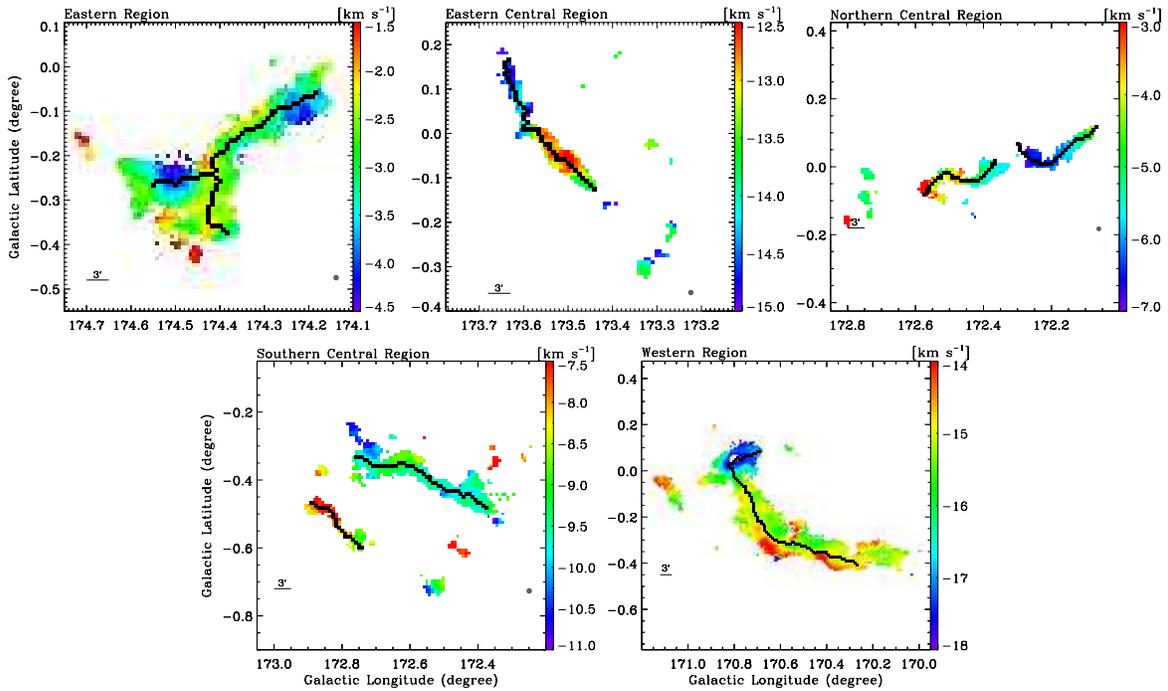}
\caption{Filamentary structures identified in each region. The solid black lines indicate the positions of the filaments identified by FilFinder. The backgrounds are the $^{13}$CO velocity distribution maps of each region.}
\label{fig:f4}
\end{figure}

In the practical identification process, we define the mask area as the emission area where the signal-to-noise ratio (S/N) is greater than 3 within the input intensity image. We set the ``adapt\_thresh'' as 0.8~pc (two times the beam size of our observations for a distance of $\sim$1.7~kpc; see Section~\ref{sec:res:pro}), which is the width of the element within the mask, and set the ``size\_thresh'' as 3.92~pc$^2$ \citep[assuming an aspect ratio of 5 according to the definition of][]{and14}, which is the minimum area a region of the mask must have. When inputting a defined mask, the algorithm will use these two parameters to verify whether this mask has qualified width and area or not. After deriving the skeletons, the algorithm will further prune the small branches attached to the skeletons with lengths smaller than the ``branch\_thresh,'' which is set as 10~pixels (five times the beam size of our observations for a pixel size of $30\arcsec \times 30\arcsec$). Based on these settings, FilFinder is applied to the $^{13}$CO intensity image of each subregion. The final result is shown in Figure~\ref{fig:f4}.

We realize that FilFinder can only identify filamentary structures in the 2D image at this point, which means it cannot take into account the velocity information. Therefore, the identified skeletons (including the attached branches) may not be velocity-coherent structures. We further inspect the $^{13}$CO data cube of each subregion and $^{13}$CO $PV$ map of each structure to make sure these skeletons and branches are velocity-coherent. Specifically, we first calculate the thermal line width of the $^{13}$CO emission area with an S/N greater than 3 within each subregion using the formula
\begin{equation}
\Delta v_{\rm thermal} = \sigma_{\rm thermal} \sqrt{8 \rm ln 2} = \sqrt{\frac{k_{\rm B} T_{\rm kin}}{\mu_{\rm obs} m_{\rm H}}} \sqrt{8 \rm ln 2}.
\end{equation}
Here $k_{\rm B}$ is the Boltzmann constant, $\mu_{\rm obs}$ is the molecular weight of the observed molecule (29 for $^{13}$CO), and $m_{\rm H}$ is the mass of the hydrogen atom. In this work, the kinetic temperature $T_{\rm kin}$ is assumed to be the excitation temperature under the local thermodynamic equilibrium (LTE) condition (see Section~\ref{sec:res:east}). We then check each two adjacent independent pixels with S/N $>$ 3 along each skeleton or branch within the data cube of each subregion. If the velocity difference of every 2 pixels is smaller than the corresponding thermal line width, we will regard this skeleton or branch as a velocity-coherent structure. Otherwise, we will reject the branch or part of the skeleton that is not velocity-coherent.

In the following subsections, we show the results of the velocity-coherence check and present the properties of filamentary structures in each subregion separately.

\subsection{Filamentary Structures in the Eastern Region}\label{sec:res:east}
\begin{figure}
\epsscale{1.0}
\plotone{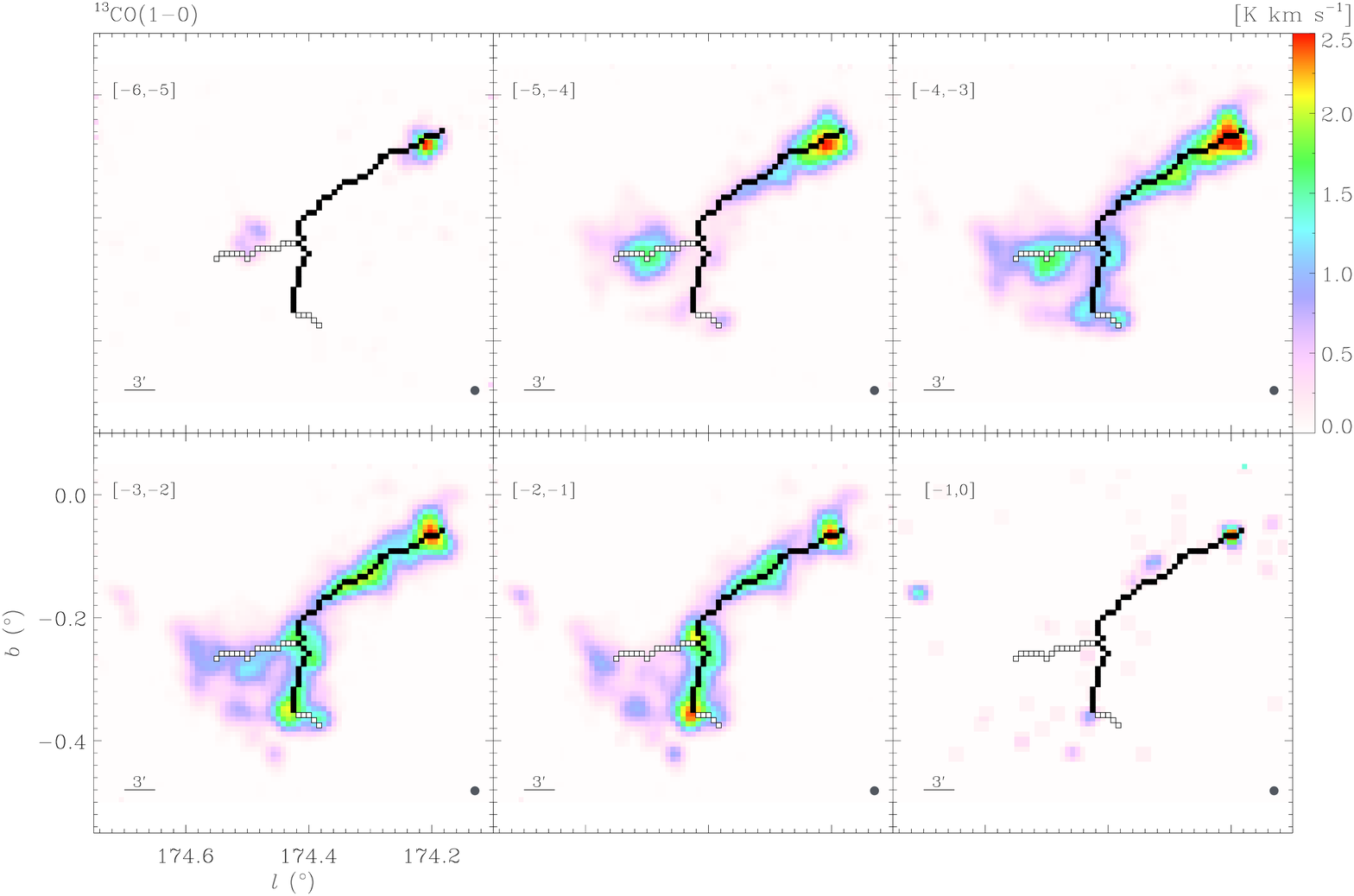}
\caption{Velocity channel maps of molecular clouds in the eastern region traced by the $^{13}$CO emission. The solid black lines indicate the qualified filament after the velocity-coherence check. The solid white lines indicate the rejected structures that are not velocity-coherent. The velocity range (in the units of km~s$^{-1}$) marked on each map is the integration range of each channel.}
\label{fig:f5}
\end{figure}

Figure~\ref{fig:f5} shows the $^{13}$CO velocity channel maps of the filamentary molecular cloud in the eastern region. The solid lines (both black and white lines) are the skeleton and branch identified by FilFinder. During the velocity-coherence check, we find that the southern end ($l \sim 174\arcdeg.4, b \sim -0\arcdeg.38$) of the skeleton is not velocity-coherent relative to the rest of the skeleton. So we reject this small part and mark it white. FilFinder has also identified a branch across a small molecular cloud ($l \sim 174\arcdeg.50\arcdeg, b \sim -0\arcdeg.26$) to the east of the skeleton. This small cloud has a different velocity component (ranging from $-5$ to $-3$~km~s$^{-1}$) from the emission area of the skeleton. Thus, we choose to reject this branch and mark it white. The black line is the qualified filamentary structure after the velocity-coherence check. We name this filamentary structure the ``eastern region filament'' (hereafter ``E-F'').

\begin{figure}
\epsscale{1.0}
\plotone{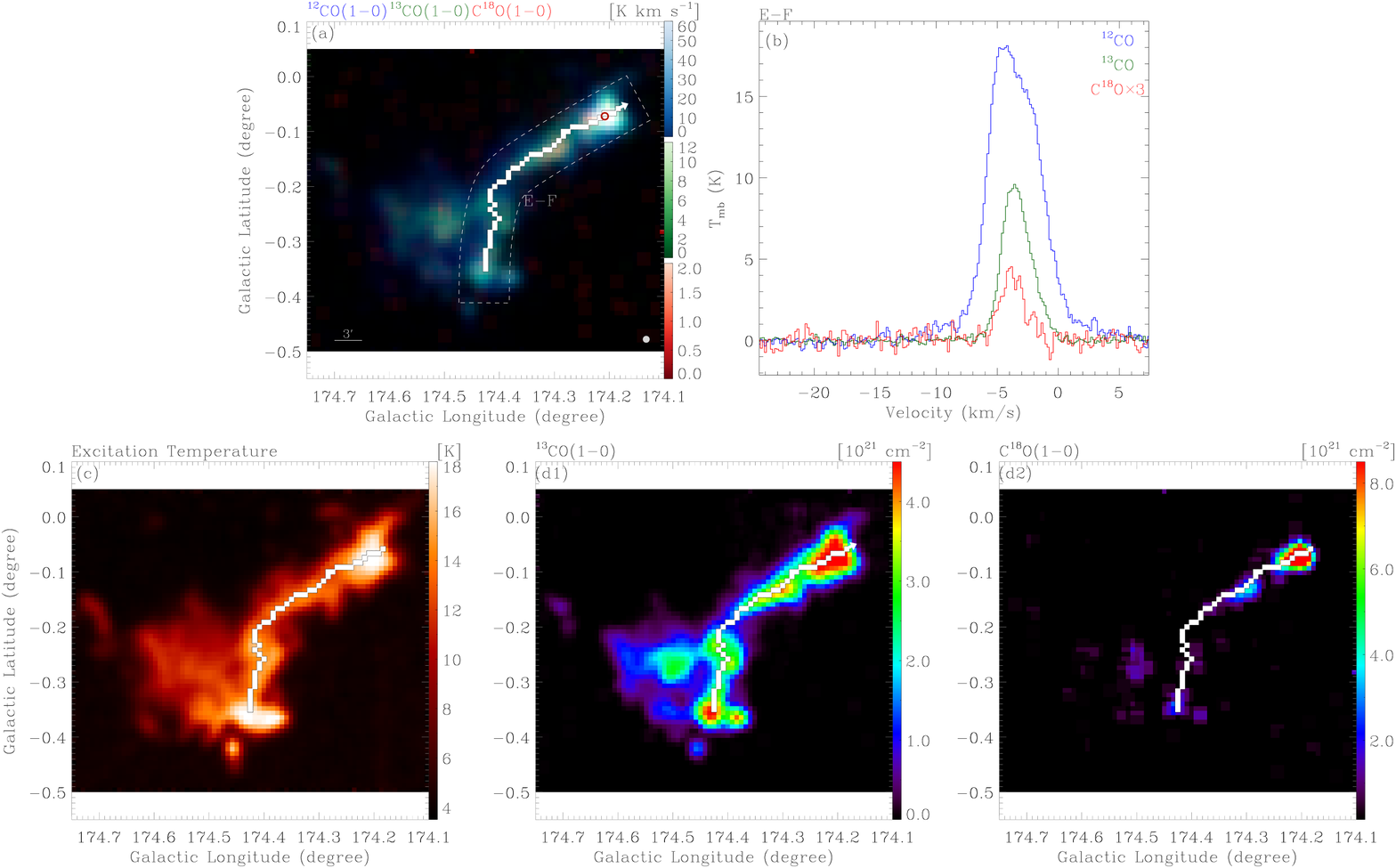}
\caption{Properties of E-F in the eastern region. (a) Three-color image of the integrated intensity maps of $^{12}$CO (blue; integrated from $-$8.0 to 1.0~km~s$^{-1}$), $^{13}$CO (green; integrated from $-$6.2 to $-$0.3~km~s$^{-1}$), and C$^{18}$O (red; integrated from $-$5.5 to $-$1.5~km~s$^{-1}$). The solid white line shows the position of E-F, and the arrow indicates the direction of the $PV$ plot of E-F shown in Figure~\ref{fig:f15} and Figure~\ref{fig:f16}. The dashed box indicates the approximate emission area around the filament where the S/N is greater than 3. The red circle indicates the peak position of the $^{12}$CO emission along the filament. (b) The CO spectra toward the red circle. (c) Excitation temperature map. (d1) and (d2) The H$_2$ column density maps traced by $^{13}$CO and C$^{18}$O. The arrow in (d1) indicates the direction used to determine the left and right sides of E-F corresponding to the $R < 0$ and $R > 0$ parts of the radial density profile in Figure~\ref{fig:f17}.}
\label{fig:f6}
\end{figure}

In panel (a) of Figure~\ref{fig:f6}, we present the three-color image of the integrated intensity maps of the filamentary structure E-F. The dashed box indicates the approximate emission area of E-F where the S/N is greater than 3. The C$^{18}$O emission appears at the brightest parts (e.g., $l \sim 174\arcdeg.21, b \sim -0\arcdeg.08$, $l \sim 174\arcdeg.30, b \sim -0\arcdeg.13$, and $l \sim 174\arcdeg.43, b \sim -0\arcdeg.34$), where the $^{12}$CO (blue), $^{13}$CO (green), and C$^{18}$O (red) emission overlap with each other. Panel (b) is the spectra of the CO isotopologues toward the red circle, which indicates the peak position of the $^{12}$CO emission.

We further calculate the excitation temperature and the column density of this filament based on the assumption that the molecular cloud is in the LTE condition, the $^{12}$CO emission is optically thick, and the $^{13}$CO emission is optically thin. To verify the CO optical depth, we adopt the correlation
\begin{equation}
\frac{T_{\rm mb}(\rm ^{12}CO)}{T_{\rm mb}(\rm ^{13}CO)} = \frac{1-e^{-\tau_{12}}}{1-e^{-\tau_{13}}} = \frac{1-e^{-\tau_{13}X}}{1-e^{-\tau_{13}}},
\end{equation}
assuming the same excitation temperature for $^{12}$CO and $^{13}$CO. Here $T_{\rm mb}$ is the main beam brightness temperature of the CO emission, $\tau$ is the optical depth of CO, and $X$ is the abundance ratio [$^{12}$CO]/[$^{13}$CO] \citep[e.g.,][]{sco86, gar91}, which is estimated to be 73.24 for the eastern region with a distance of 1.68~kpc (see Section~\ref{sec:res:pro}) based on the work of \citet{mil05}. The result is that the optical depth of $^{12}$CO ($\tau_{12}$) ranges from 3.22 to 46.87 within the velocity range of filament E-F, and the optical depth of $^{13}$CO ($\tau_{13}$) ranges from 0.04 to 0.64. We also verify the CO optical depth for the rest of the subregions and find that $\tau_{12}$ is much larger than 1 and $\tau_{13}$ is smaller than 1 within the velocity components of the filamentary molecular clouds.

After the verification of the CO optical depth, we can calculate the excitation temperature ($T_{\rm ex}$) from the brightness temperature of the $^{12}$CO emission using 
\begin{equation}
T_{\rm ex} = T_0 / \ln \left( 1 + \left( \frac{T_{\rm mb}(\rm ^{12}CO)}{(1-e^{-\tau_{12}})T_0} + \frac{1}{e^{T_0/T_{\rm bg}}-1} \right)^{-1} \right).
\end{equation}
Here $T_0 = h\nu/k_{\rm B}$ is the intrinsic temperature of $^{12}$CO, where $h$ is the Planck constant and $k_{\rm B}$ is the Boltzmann constant, and $T_{\rm bg}$ is the background temperature with a value of 2.7~K. The opacity depth $\tau_{12}$ is much larger than 1, which means $1-e^{-\tau_{12}} \approx 1$. Under all of these conditions, we derive the excitation temperature map and present it in panel (c) of Figure~\ref{fig:f6}. The temperature is much higher at the two ends of E-F ($\sim$21~K in the northern end and $\sim$23~K in the southern end) compared to the middle part. The mean temperature of the whole filament is about 8.72$\pm$0.97~K.

In this work, we choose to regard the excitation temperature ($T_{\rm ex}$) derived from the $^{12}$CO emission as the kinetic temperature ($T_{\rm kin}$) of the molecular gas. The larger optical depth of $^{12}$CO ($\tau_{12} \gg 1$) increase the interactions between the photons and the matter, which tends to lead to the LTE condition. We further check the dust temperature ($T_{\rm dust}$) map derived from $Herschel$ for the eastern region \citep{mar17}. \citet{mar17} processed the Hi-GAL data with the point process mapping (PPMAP) method \citep{mar15}, which is a Bayesian procedure that uses images of dust continuum emission at multiple wavelengths and the associated point-spread functions (PSFs) to produce resolution-enhanced image cubes of column density and dust temperature. The dust temperature has a similar distribution as the excitation temperature, increasing toward the denser parts of the filamentary structure. The mean ratio of $T_{\rm dust}$ to $T_{\rm ex}$ within the filamentary area is about 1.0239. Based on the LTE condition and the similar distribution as the dust temperature, it is reasonable to use $T_{\rm ex}$ as $T_{\rm kin}$ in this work, since the $Herschel$ observations \citep{mar17} do not cover the entire area of the G172 region.

We also derive the H$_2$ column density maps of E-F traced by the $^{13}$CO and C$^{18}$O emission, which are shown in the bottom panels of Figure~\ref{fig:f6}. Since the $^{13}$CO and C$^{18}$O emission is optically thin in the velocity component of E-F, we can first calculate the column density using the excitation temperature and their integrated intensity, then multiply the column density by the abundance of $N_{\rm H_2}/N_{\rm ^{13}CO}$ \citep[7$\times$10$^{5}$;][]{fre82} and $N_{\rm H_2}/N_{\rm C^{18}O}$ \citep[7$\times$10$^{6}$;][]{cas95} to derive the H$_2$ column density \citep[see the details in][]{xio17}. The mean values of H$_2$ column density are 4.77$\pm$0.49~$\times$10$^{21}$~cm$^{-2}$ (within the $^{13}$CO emission area with S/N $>$ 3) and 8.37$\pm$0.11~$\times$10$^{21}$~cm$^{-2}$ (within the C$^{18}$O emission area with S/N $>$ 3).

We further compare the column density derived from the $^{13}$CO data with that derived from the far-IR (FIR) data obtained with $Herschel$ for the eastern region and eastern central regions \citep{mar17}. For the eastern region, the mean ratio of $N_{\rm H_2}$ ($^{13}$CO) to $N_{\rm H_2}$ (FIR) is about 0.45, and for the eastern central region, the mean ratio is about 0.38. The H$_2$ column density from the FIR emission is typically about two to three times higher than the H$_2$ column density measured by the $^{13}$CO emission. However, \citet{san17} pointed out that the mass estimation of molecular gas from dust emission could have an uncertainty higher than a factor of 2, resulting from the large uncertainties of the dust opacity and the dust-to-gas mass ratio. Indeed, as mentioned by \citet{mar17}, the reference dust opacity at 300~$\mu$m could have the largest uncertainty up to about 50$\%$. The dust-to-gas mass ratio of 100 \citep[adapted by][]{mar17} could also have an uncertainty of 23$\%$ \citep{san17}, assuming the uniform distribution of the range of the Galactic dust-to-gas mass ratio between 70 and 150 \citep[e.g.,][]{dev90, vuo03}. The final uncertainties of the H$_2$ column density derived by \citet{mar17} have typical values as high as a factor of 1.8. Given these similar factors of uncertainty values, the agreement between the column density estimations from the CO and FIR data is reasonable. Therefore, in this work, it is reasonable to use the CO emission to calculate the H$_2$ column density.

With the direction of the arrowed line and the width of the dashed box shown in panel (a) of Figure~\ref{fig:f6}, we extract the $^{13}$CO and C$^{18}$O position-velocity ($PV$) plots of E-F (see Figure~\ref{fig:f15} and Figure~\ref{fig:f16}, respectively). The $^{13}$CO $PV$ plot displays continuous structure without significant curvature along the position axis. The velocity gradient of E-F is weak, which is from $-$2.5~km~s$^{-1}$ (southern part) to $-$3.5~km~s$^{-1}$ (northwestern part; referring to the direction of E-F in Figure~\ref{fig:f6}). Tracing the densest parts of E-F, the C$^{18}$O $PV$ plot displays discontinuous features. A strong emission area is located in the northwest, while several weaker emission spots are in the south.

\subsection{Filamentary Structures in the Eastern Central Region}
\begin{figure}
\epsscale{1.0}
\plotone{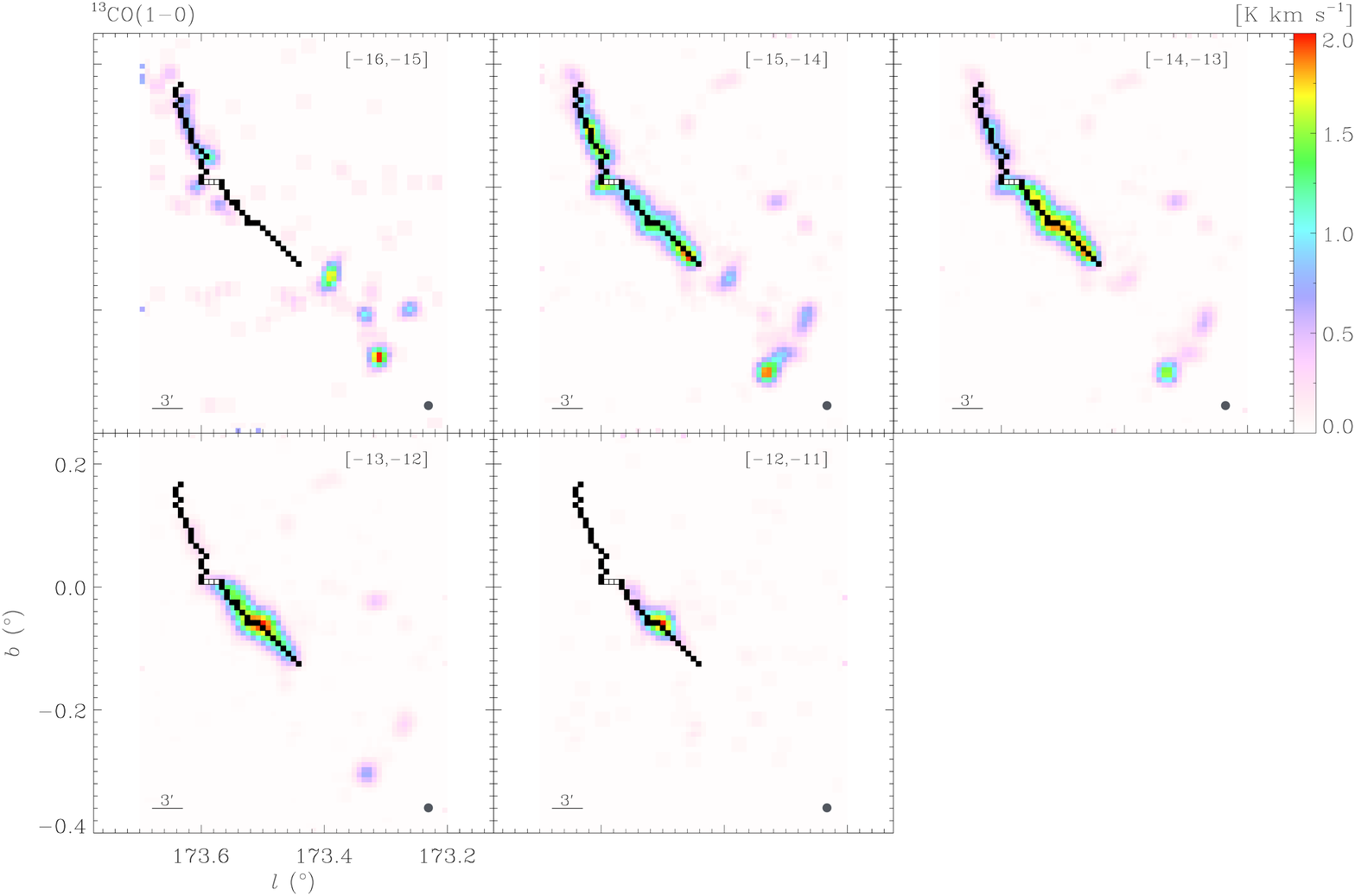}
\caption{Velocity channel maps of molecular clouds in the eastern central region traced by the $^{13}$CO emission. The solid black lines indicate the qualified filaments after the velocity-coherence check. The solid white lines indicate the rejected structure that is not velocity-coherent. The velocity range (in the units of km~s$^{-1}$) marked on each map is the integration range of each channel.}
\label{fig:f7}
\end{figure}

In the eastern central region, FilFinder has identified only one skeleton (see both the black and white solid lines in Figure~\ref{fig:f7}). However, during the velocity-coherence check, we find that the northern part of the skeleton and the southern part are not velocity-coherent relative to each other. Actually, the velocity component (ranging from $-15.5$ to $-13.8$~km~s$^{-1}$) of the northern part is different from that of the southern part (ranging from $-13.8$ to $-11.2$~km~s$^{-1}$). So we decide to split this long skeleton into two skeletons by rejecting a small middle part (marked as white in Figure~\ref{fig:f7}) of the long skeleton. The black lines in Figure~\ref{fig:f7} show the qualified filamentary structures. We name the northern filament the ``eastern central region filament (north)'' (hereafter ``EC-F(N)'') and the southern filament the ``eastern central region filament (south)'' (hereafter ``EC-F(S)''). Both of these two filaments extend from the northeast to the southwest.

The integrated intensity maps of the CO isotopologues are shown in panel (a) of Figure~\ref{fig:f8}. The dashed boxes indicate the approximate emission areas of EC-F(S) and EC-F(N) where the S/N is greater than 3. We detect stronger $^{13}$CO and C$^{18}$O emission in EC-F(S), while in EC-F(N), the $^{13}$CO emission is much weaker, and no C$^{18}$O emission is detected (S/N $<$ 3).

The CO spectra in panels (b1) and (b2) show the differences in emission intensities and velocity components between these two filaments as we discussed above. The excitation temperature and column density of EC-F(S) are higher than those of EC-F(N) (see panels (c), (d1), and (d2)). The mean $T_{\rm ex}$ values are 8.02$\pm$0.82~K (EC-F(N)) and 9.60$\pm$1.23~K (EC-F(S)), and the mean $N_{\rm H_2}$ values are 1.26$\pm$0.16~$\times$10$^{21}$~cm$^{-2}$ ($^{13}$CO) for EC-F(N) and 6.53$\pm$0.69~$\times$10$^{21}$~cm$^{-2}$ ($^{13}$CO) and 7.94$\pm$0.12~$\times$10$^{21}$~cm$^{-2}$ (C$^{18}$O) for EC-F(S).

The $^{13}$CO $PV$ plots are presented in Figure~\ref{fig:f15}. Both EC-F(N) and EC-F(S) have very weak velocity gradients. Here EC-F(N) has several strong emission spots along the position axis, while EC-F(S) has one strong emission spot located in the central part. The C$^{18}$O $PV$ plot of EC-F(S) is shown in Figure~\ref{fig:f16}; the C$^{18}$O emission is only detected in the central part.

\subsection{Filamentary Structures in the Northern Central Region}
\begin{figure}
\epsscale{1.0}
\plotone{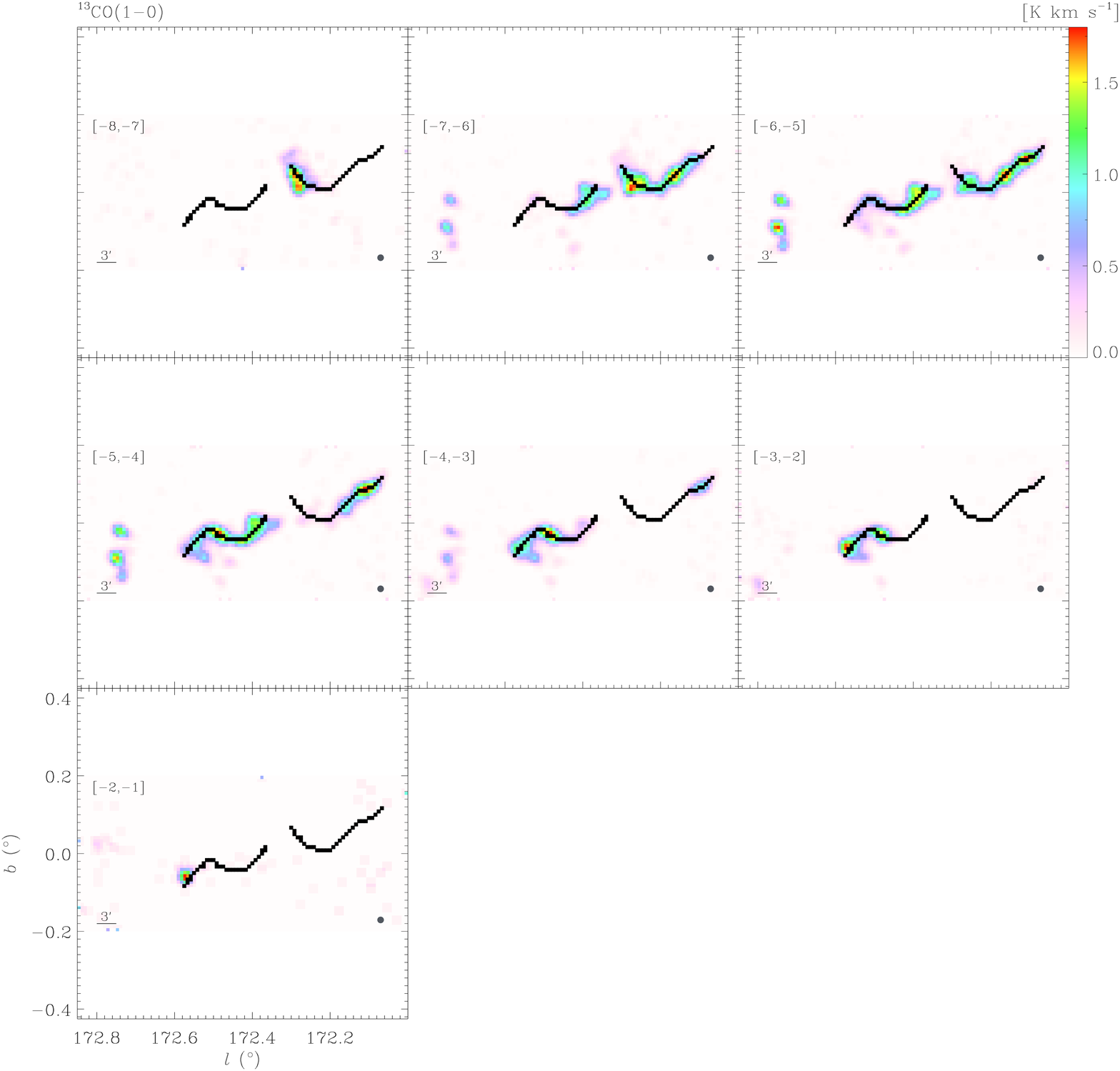}
\caption{Velocity channel maps of molecular clouds in the northern central region traced by the $^{13}$CO emission. The solid black lines indicate the qualified filaments after the velocity-coherence check. The velocity range (in the units of km~s$^{-1}$) marked on each map is the integration range of each channel.}
\label{fig:f9}
\end{figure}

Shown in Figure~\ref{fig:f9}, FilFinder has identified two skeletons in the northern central region. During the velocity-coherence check, both of these skeletons showed velocity-coherent structures. So we regard them as qualified filaments. The filament in the east (``northern central region filament (east),'' hereafter ``NC-F(E)'') has a velocity component ranging from $-7$ to $-2$~km~s$^{-1}$, while the filament in the west (``northern central region filament (west),'' hereafter ``NC-F(W)'') is distributed from $-8$ to $-4$~km~s$^{-1}$. Both of these filaments exhibit curvy shapes extending in the east--west direction.

\begin{figure}
\epsscale{1.0}
\plotone{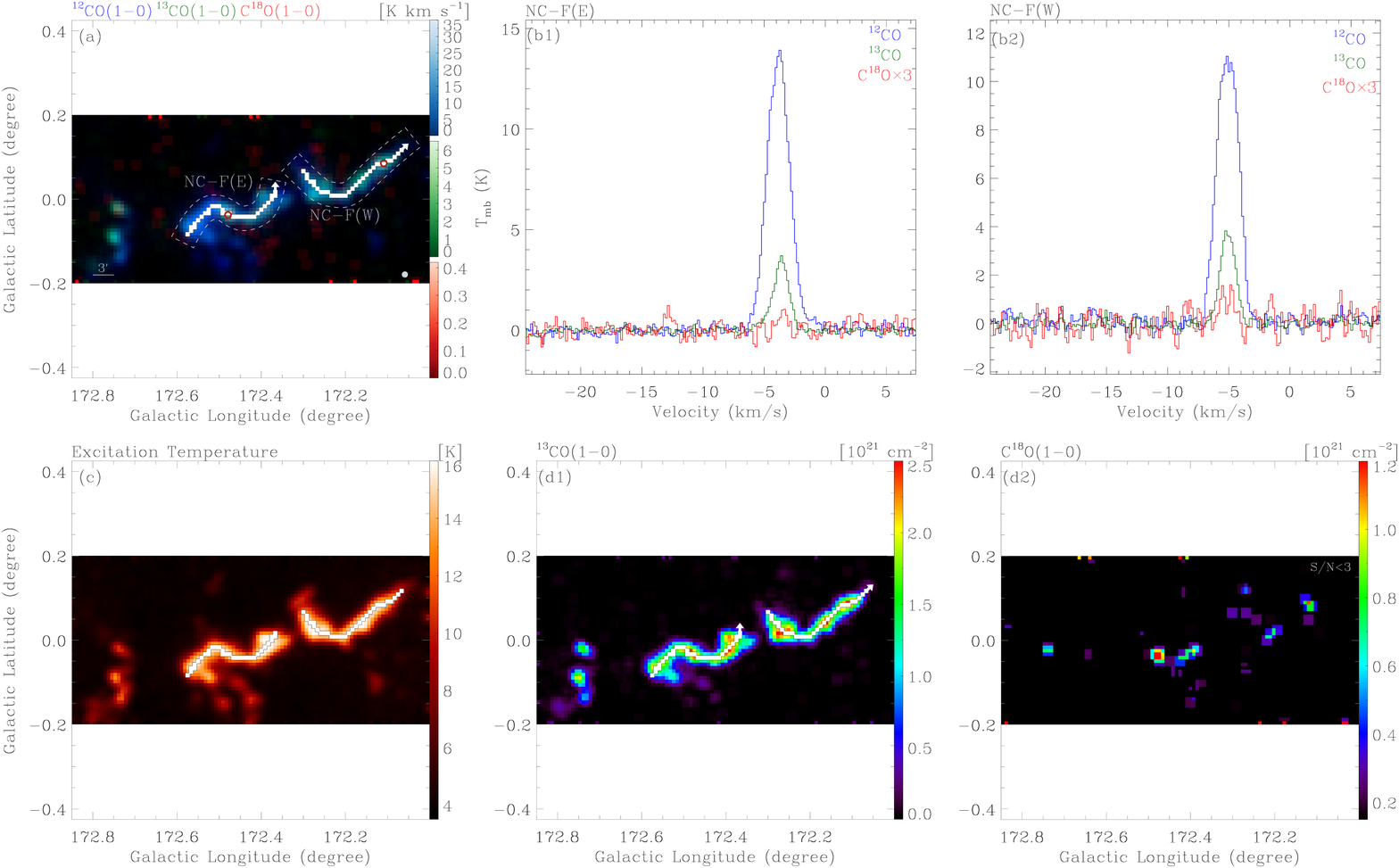}
\caption{Properties of NC-F(E) and NC-F(W) in the northern central region. (a) Three-color image of theintegrated intensity maps of $^{12}$CO (blue; integrated from $-$9.0 to 0.5~km~s$^{-1}$), $^{13}$CO (green; integrated from $-$8.2 to $-$0.7~km~s$^{-1}$), and C$^{18}$O (red; integrated from $-$5.9 to $-$3.1~km~s$^{-1}$). The solid white lines show the positions of NC-F(E) and NC-F(W), and the arrows indicate the directions of the $PV$ plots of NC-F(E) and NC-F(W) shown in Figure~\ref{fig:f15}. The dashed boxes indicate the approximate emission area around each filament where the S/N is greater than 3. The red circles indicate the peak positions of the $^{12}$CO emission along each filament. (b1) and (b2) The CO spectra toward the red circles. (c) Excitation temperature map. (d1) and (d2) The H$_2$ column density maps traced by $^{13}$CO and C$^{18}$O. The arrows in (d1) indicate the directions used to determine the left and right sides of NC-F(E) and NC-F(W) corresponding to the $R < 0$ and $R > 0$ parts of the radial density profiles in Figure~\ref{fig:f17}.}
\label{fig:f10}
\end{figure}

Figure~\ref{fig:f10} shows the basic properties of NC-F(E) and NC-F(W). The $^{13}$CO emission of NC-F(W) is slightly stronger than that of NC-F(E), and no C$^{18}$O emission is detected (S/N $<$ 3) in both filaments (see panels (a), (b1), and (b2)). The excitation temperatures are similar between these two filaments: 8.93$\pm$0.96~K (NC-F(E)) and 9.41$\pm$1.01~K (NC-F(W)). The mean H$_2$ column density is 1.53$\pm$0.19~$\times$10$^{21}$~cm$^{-2}$ ($^{13}$CO) for NC-F(E), while for NC-F(W), the value is 1.72$\pm$0.23~$\times$10$^{21}$~cm$^{-2}$ ($^{13}$CO).

Shown in Figure~\ref{fig:f15}, the $^{13}$CO $PV$ plot of NC-F(E) displays a structure consisting of three different parts. The southeastern and middle parts appear to be split from each other but have the same velocity. The northwestern part is connected to the middle part but has a different velocity. The velocity gradient of NC-F(E) is from $-$3.0~km~s$^{-1}$ (southeastern part; referring to the direction of NC-F(E) in Figure~\ref{fig:f10}) to $-$5.5~km~s$^{-1}$ (northwestern part), while the velocity gradient of NC-F(W) is from $-$7.0~km~s$^{-1}$ (northeastern part; referring to the direction of NC-F(W) in Figure~\ref{fig:f10}) to $-$4.5~km~s$^{-1}$ (northwestern part).

\subsection{Filamentary Structures in the Southern Central Region}
\begin{figure}
\epsscale{1.0}
\plotone{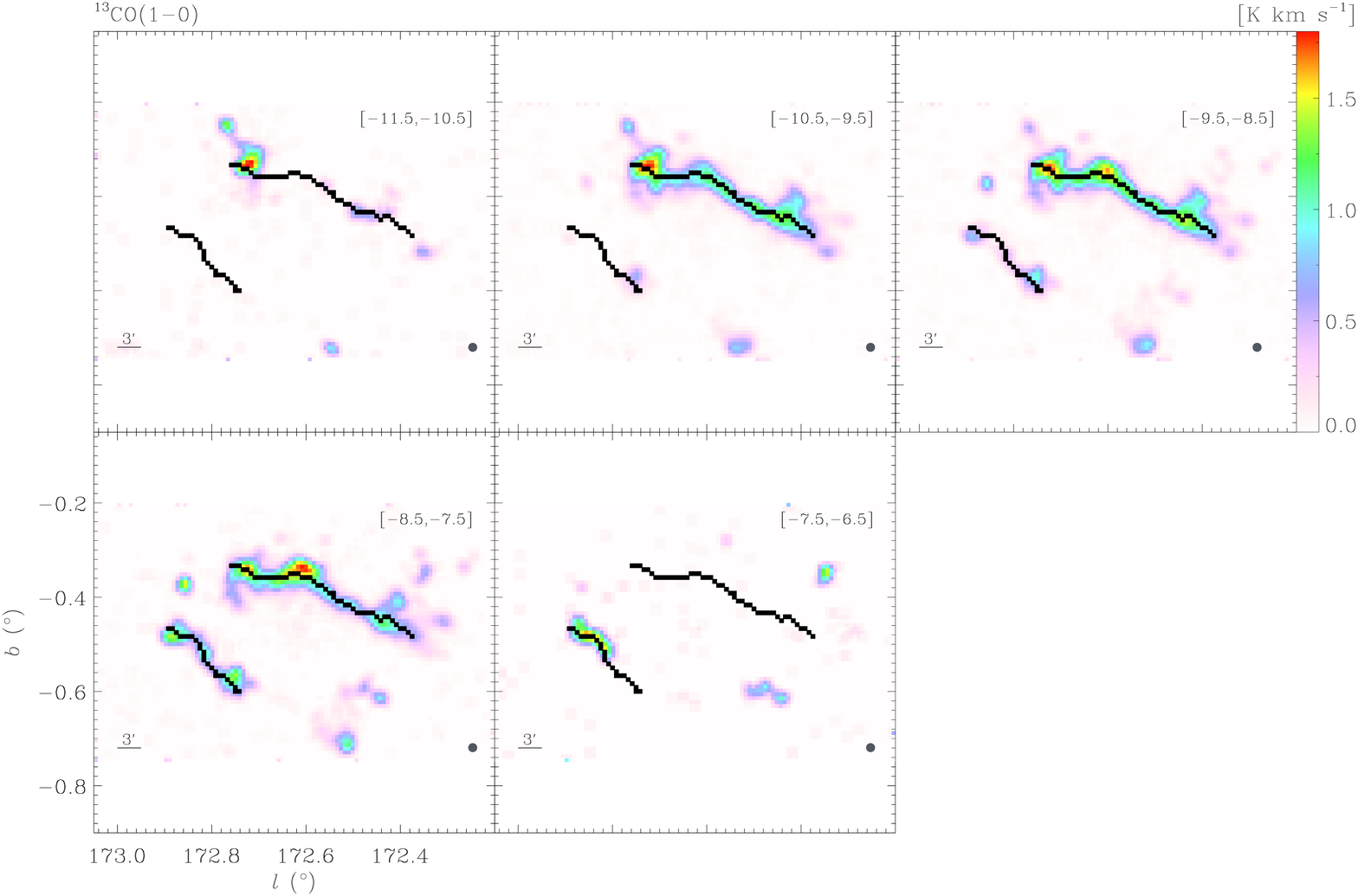}
\caption{Velocity channel maps of molecular clouds in the southern central region traced by the $^{13}$CO emission. The solid black lines indicate the qualified filaments after the velocity-coherence check. The velocity range (in the units of km~s$^{-1}$) marked on each map is the integration range of each channel.}
\label{fig:f11}
\end{figure}

Two individual skeletons are identified in this region (see Figure~\ref{fig:f11}). These two skeletons are velocity-coherent and thus qualified filaments after the velocity-coherence check. We name the northern one ``SC-F(N),'' for which the velocity ranges from $-11$ to $-8.5$~km~s$^{-1}$. The southern one is named ``SC-F(S)'' and is revealed from $-9.5$ to $-7$ km~s$^{-1}$. These two filaments are parallel with each other, and SC-F(N) extends much farther than SC-F(S).

Shown in panels (a), (b1), and (b2) of Figure~\ref{fig:f12}, SC-F(N) has much stronger $^{12}$CO and $^{13}$CO emission than SC-F(S), and the C$^{18}$O emission is only detected in SC-F(N). The mean excitation temperature is about 8.99$\pm$0.99~K in SC-F(N) and lower in SC-F(S) (6.16$\pm$0.49~K). The bottom panels present the H$_2$ column density maps of these two filaments. For SC-F(N), the mean column densities are 2.94$\pm$0.32~$\times$10$^{21}$~cm$^{-2}$ ($^{13}$CO) and 5.05$\pm$0.07~$\times$10$^{21}$~cm$^{-2}$ (C$^{18}$O). For SC-F(S), the mean value is 0.70$\pm$0.09~$\times$10$^{21}$~cm$^{-2}$ ($^{13}$CO).

Both SC-F(N) and SC-F(S) have slightly twisted velocity structures (see Figure~\ref{fig:f15}). The velocity component of SC-F(N) changes from $-$9.5~km~s$^{-1}$ (western part) to $-$9.0~km~s$^{-1}$ (middle part) and then changes to $-$10.0~km~s$^{-1}$ (eastern part). A similar velocity gradient is found in SC-F(S), which goes from $-$9.0~km~s$^{-1}$ (western part) to $-$7.5~km~s$^{-1}$ (middle part) and changes to $-$8.0~km~s$^{-1}$ (eastern part). The C$^{18}$O $PV$ plot of SC-F(N) is shown in the last panel of Figure~\ref{fig:f16}. The emission area is mainly located in the eastern part.

\subsection{Filamentary Structures in the Western Region}
\begin{figure}
\epsscale{1.0}
\plotone{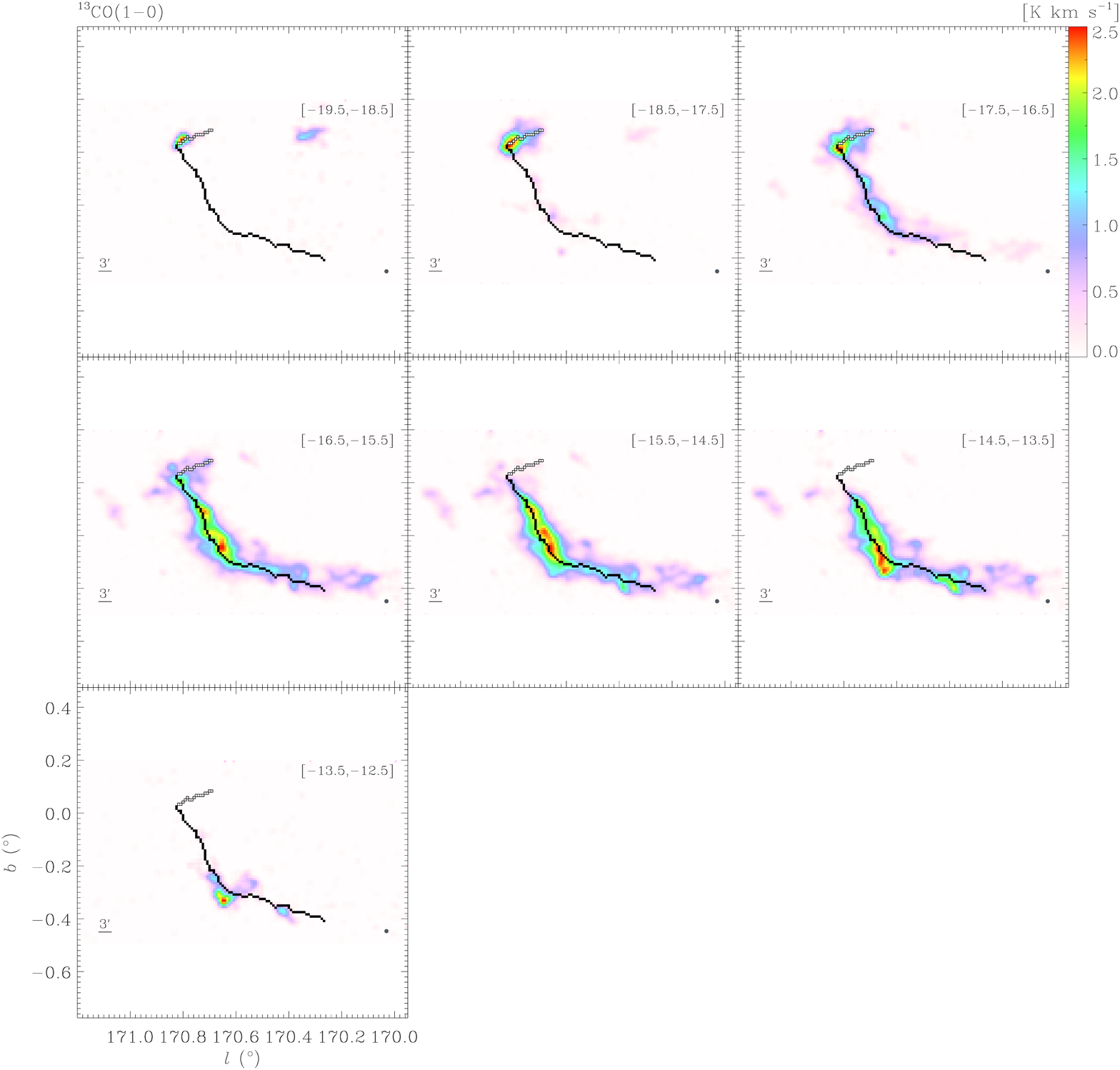}
\caption{Velocity channel maps of molecular clouds in the western region traced by the $^{13}$CO emission. The solid black lines indicate the qualified filament after the velocity-coherence check. The solid white lines indicate the rejected structure that is not velocity-coherent. The velocity range (in the units of km~s$^{-1}$) marked on each map is the integration range of each channel.}
\label{fig:f13}
\end{figure}

The velocity channel maps of the western region are presented in Figure~\ref{fig:f13}. Only one long skeleton is identified by FilFinder. During the velocity-coherence check, we find that the northeastern end ($l \sim 170\arcdeg.82, b \sim 0\arcdeg.02$) of the long skeleton has a different velocity component (ranging from $-19.5$ to $-17.5$~km~s$^{-1}$) from the main body of the skeleton (ranging from $-17$ to $-13.5$~km~s$^{-1}$) and choose to reject this part (see the white line in Figure~\ref{fig:f13}). The qualified filament is marked by a black line in Figure~\ref{fig:f13}. We name this filament the ``western region filament'' (``W-F''). The whole filament extends from northeast to southwest.

\begin{figure}
\epsscale{1.0}
\plotone{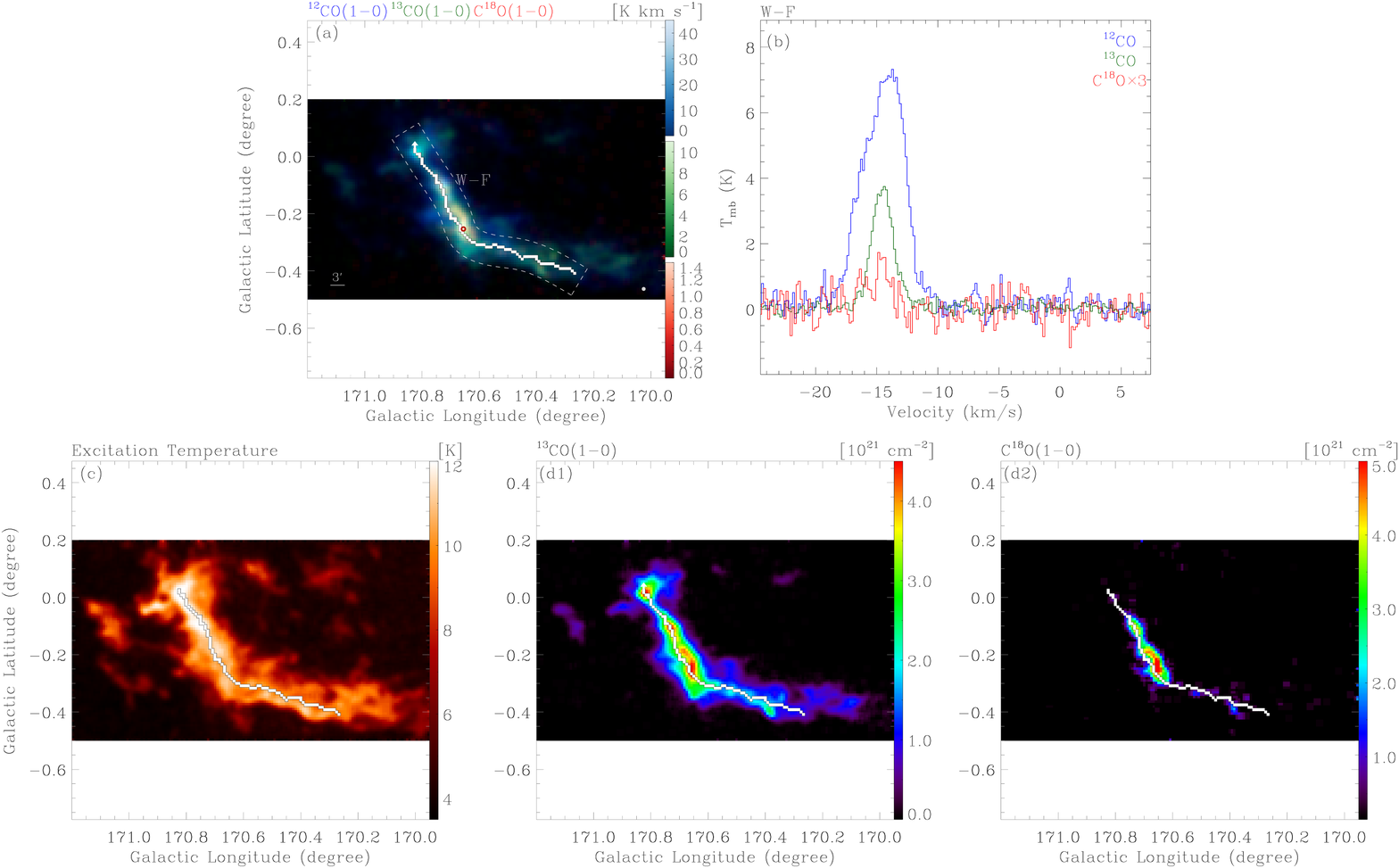}
\caption{Properties of W-F in the Western region. (a) Three-color image of integrated intensity maps of $^{12}$CO (blue; integrated from $-$20.0 to $-$11.0~km~s$^{-1}$), $^{13}$CO (green; integrated from $-$19.5 to $-$12.4~km~s$^{-1}$), and C$^{18}$O (red; integrated from $-$16.8 to $-$14.1~km~s$^{-1}$). The solid white line shows the position of W-F, and the arrow indicates the direction of the $PV$ plot of W-F shown in Figure~\ref{fig:f15} and Figure~\ref{fig:f16}. The dashed box indicates the approximate emission area around the filament where the S/N is greater than 3. The red circle indicates the peak position of the $^{12}$CO emission along the filament. (b) The CO spectra toward the red circle. (c) Excitation temperature map. (d1) and (d2) The H$_2$ column density maps traced by $^{13}$CO and C$^{18}$O. The arrow in (d1) indicates the direction used to determine the left and right sides of W-F corresponding to the $R < 0$ and $R > 0$ parts of the radial density profile in Figure~\ref{fig:f17}.}
\label{fig:f14}
\end{figure}

Shown in the panel (a) of Figure~\ref{fig:f14} is the three-color image of W-F. The C$^{18}$O emission is mainly distributed in the middle part of W-F. The northern part of W-F has a higher excitation temperature than the southern part, according to the $T_{\rm ex}$ map shown in panel (c). The mean $T_{\rm ex}$ of the whole filament is about 8.13$\pm$0.74~K. We further derive the H$_2$ column density maps of W-F traced by CO isotopologues (see panels (d1) and (d2)). The mean values are 2.12$\pm$0.25~$\times$10$^{21}$~cm$^{-2}$ ($^{13}$CO) and 3.52$\pm$0.05~$\times$10$^{21}$~cm$^{-2}$ (C$^{18}$O).

For the most part, from southwest to northeast of W-F, there is no evident velocity gradient revealed (see Figure~\ref{fig:f15}). The velocity gradient occurs at the northernmost part of W-F, which is from $-$15.0~km~s$^{-1}$ to $-$17.5~km~s$^{-1}$. The C$^{18}$O $PV$ plot is presented in Figure~\ref{fig:f16}. There is a large area of C$^{18}$O emission located in the middle part, while several weaker emission spots are in the southwest and north.

\subsection{Properties of Filaments}\label{sec:res:pro}

\subsubsection{Distances to filaments}
In this work, we adopt the probability density function (PDF) generated by the Bayesian approach \citep{rei16} to estimate the distance to each filament. For a source given with its Galactic longitude and latitude coordinate and local standard-of-rest velocity ($l, b ,v$), the Bayesian distance estimation program \footnote{\url{http://bessel.vlbi-astrometry.org/bayesian}} of \citet{rei16} will construct a PDF for each type of distance information (including spiral arm assignment, kinematic distance, Galactic latitude, and location within a giant molecular cloud with a measured parallax) and then multiply them together to arrive at a combined distance PDF. The distance to this source is determined from the component fitted to the combined PDF that has the greatest integrated probability.

In our case, we first choose three points located at the two ends and in the middle of each filament and then use their coordinates ($l, b ,v$) to calculate the distances to these points. As a result, we derive three distance values in the eastern and western regions, and six distance values in the eastern central, northern central, and southern central regions. Shown in Table~\ref{tab:tab1}, the distances to the points within the same region are basically the same. So we regard the average of the distances to the points in each region as the distance to filamentary structures in each region, which are 1.68~kpc for the eastern region, 1.72~kpc for the eastern central region, 1.70~kpc for the northern central region, 1.73~kpc for the southern central region, and 1.90~kpc for the western region.

\subsubsection{LTE mass of filaments}
With the distance to each filament and H$_2$ column density derived in the previous sections, we are able to calculate the LTE mass of the filaments by
\begin{equation}
M = \mu_{\rm H_2} m_{\rm H} \int N_{\rm H_2} dS, \label{equa:mass}
\end{equation}
where $\mu_{\rm H_2}$ is the mean molecular weight per hydrogen molecule with a value of 2.83 \citep{kau08} and $S$ is the area of CO emission. Here we use the H$_2$ column density traced by $^{13}$CO as $N_{\rm H_2}$, since the $^{12}$CO emission is too diffuse and not all of the filaments are detected with the C$^{18}$O emission. We also measure the length of each filament and derive the LTE line masses $M_{\rm LTE}$, which are listed in Table~\ref{tab:tab2}.

\subsubsection{Radial profiles of filaments}
\begin{figure}
\epsscale{1.0}
\plotone{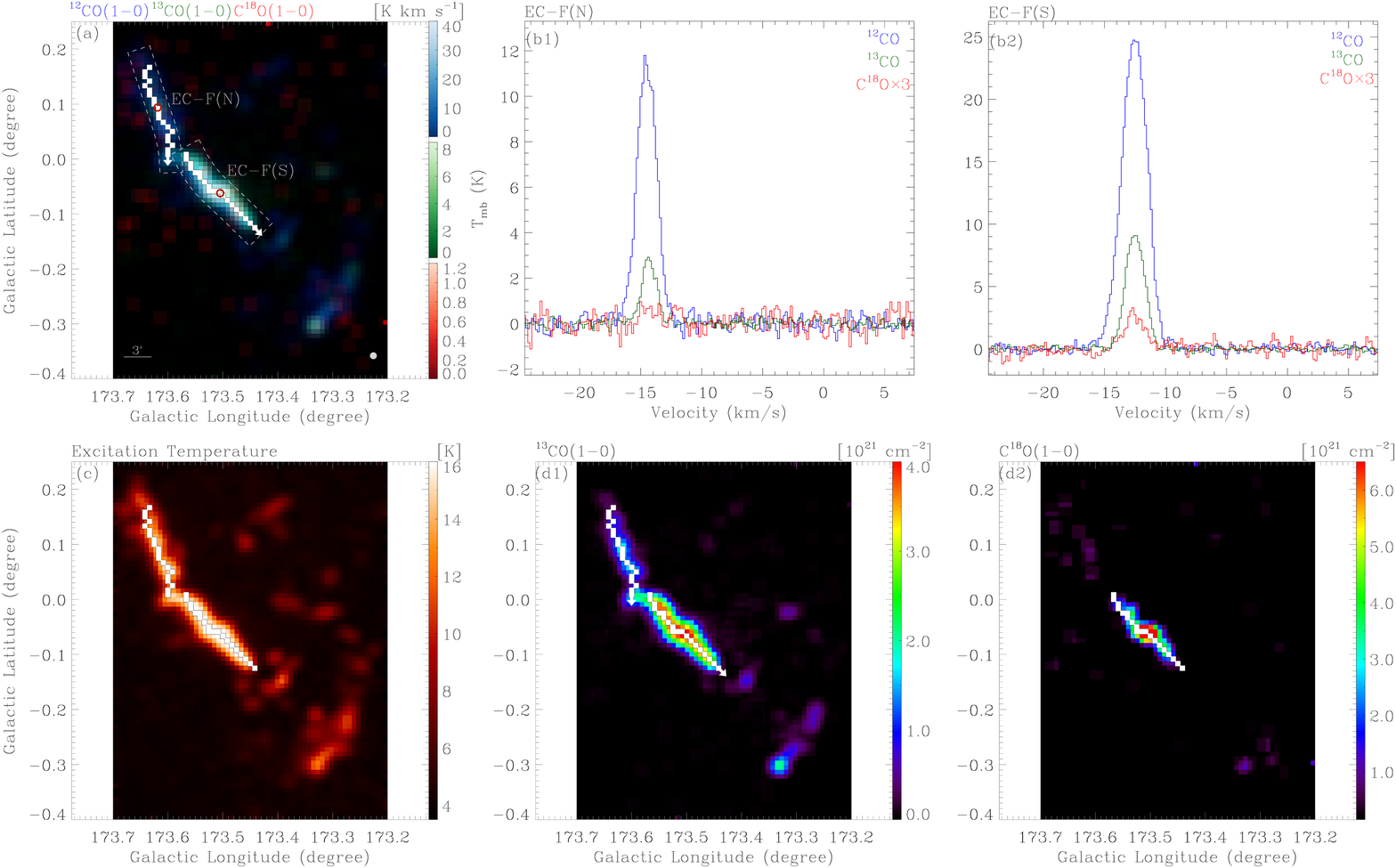}
\caption{Properties of EC-F(N) and EC-F(S) in the eastern central region. (a) Three-color image of the integrated intensity maps of $^{12}$CO (blue; integrated from $-$17.0 to $-$10.0~km~s$^{-1}$), $^{13}$CO (green; integrated from $-$15.8 to $-$10.7~km~s$^{-1}$), and C$^{18}$O (red; integrated from $-$13.9 to $-$11.3~km~s$^{-1}$). The solid white lines show the positions of EC-F(N) and EC-F(S), and the arrows indicate the directions of the $PV$ plots of EC-F(N) and EC-F(S) shown in Figure~\ref{fig:f15} and Figure~\ref{fig:f16}. The dashed boxes indicate the approximate emission area around each filament where the S/N is greater than 3. The red circles indicate the peak positions of the $^{12}$CO emission along each filament. (b1) and (b2) The CO spectra toward the red circles. (c) Excitation temperature map. (d1) and (d2) The H$_2$ column density maps traced by $^{13}$CO and C$^{18}$O. The arrows in (d1) indicate the directions used to determine the left and right sides of EC-F(N) and EC-F(S) corresponding to the $R < 0$ and $R > 0$ parts of the radial density profiles in Figure~\ref{fig:f17}.}
\label{fig:f8}
\end{figure}

\begin{figure}
\epsscale{1.0}
\plotone{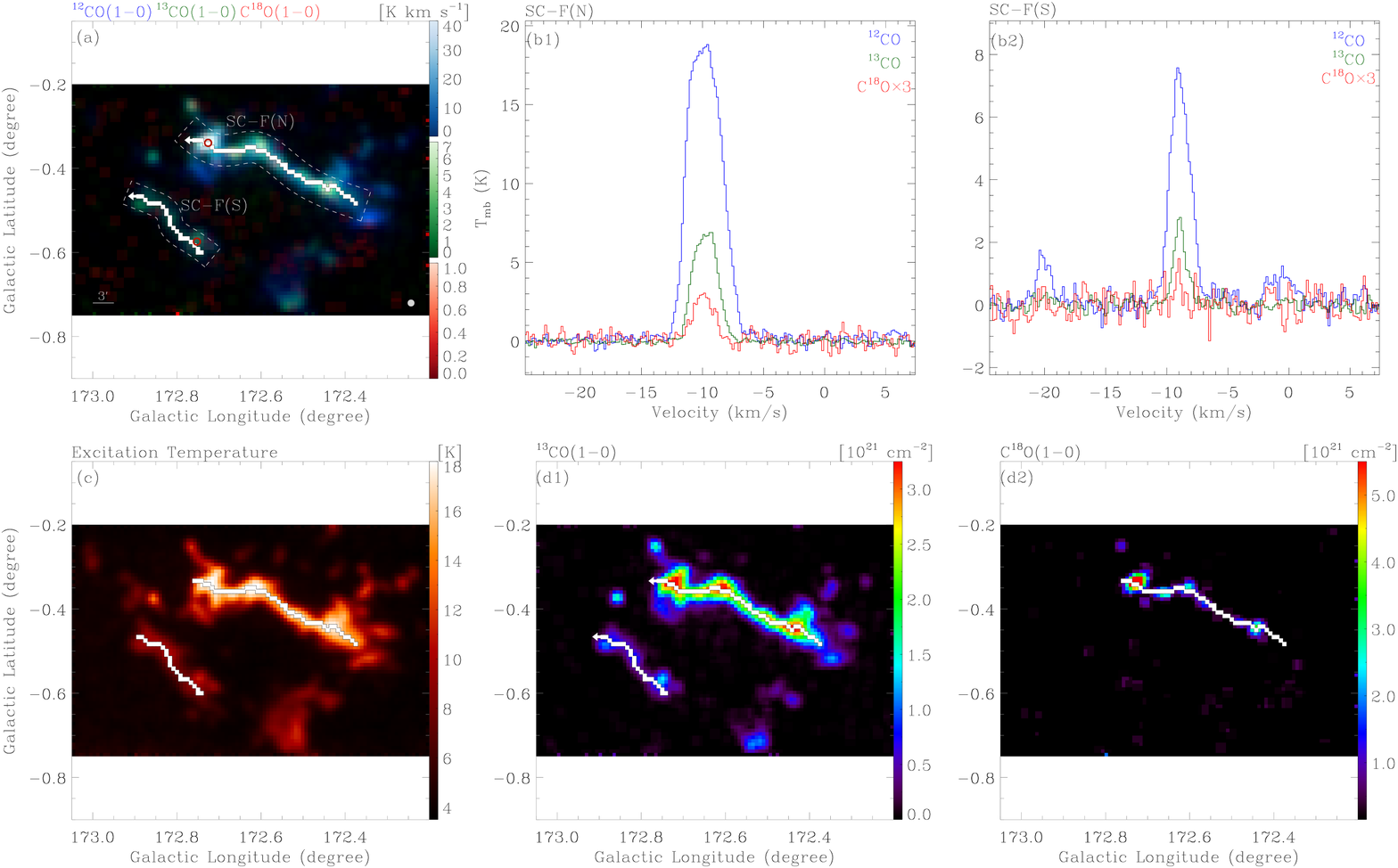}
\caption{Properties of SC-F(N) and SC-F(S) in the southern central region. (a) Three-color image of integrated intensity maps of $^{12}$CO (blue; integrated from $-$13.0 to $-$6.0~km~s$^{-1}$), $^{13}$CO (green; integrated from $-$11.9 to $-$6.4~km~s$^{-1}$), and C$^{18}$O (red; integrated from $-$11.1 to $-$8.8~km~s$^{-1}$). The solid white lines show the positions of SC-F(N) and SC-F(S), and the arrows indicate the directions of $PV$ plots of SC-F(N) and SC-F(S) shown in Figure~\ref{fig:f15} and Figure~\ref{fig:f16}. The dashed boxes indicate the approximate emission area around each filament where the S/N is greater than 3. The red circles indicate the peak positions of the $^{12}$CO emission along each filament. (b1) and (b2) The CO spectra toward the red circles. (c) Excitation temperature map. (d1) and (d2) The H$_2$ column density maps traced by $^{13}$CO and C$^{18}$O. The arrows in (d1) indicate the directions used to determine the left and right sides of SC-F(N) and SC-F(S) corresponding to the $R < 0$ and $R > 0$ parts of the radial density profiles in Figure~\ref{fig:f17}.}
\label{fig:f12}
\end{figure}

\begin{figure}
\epsscale{1.0}
\plotone{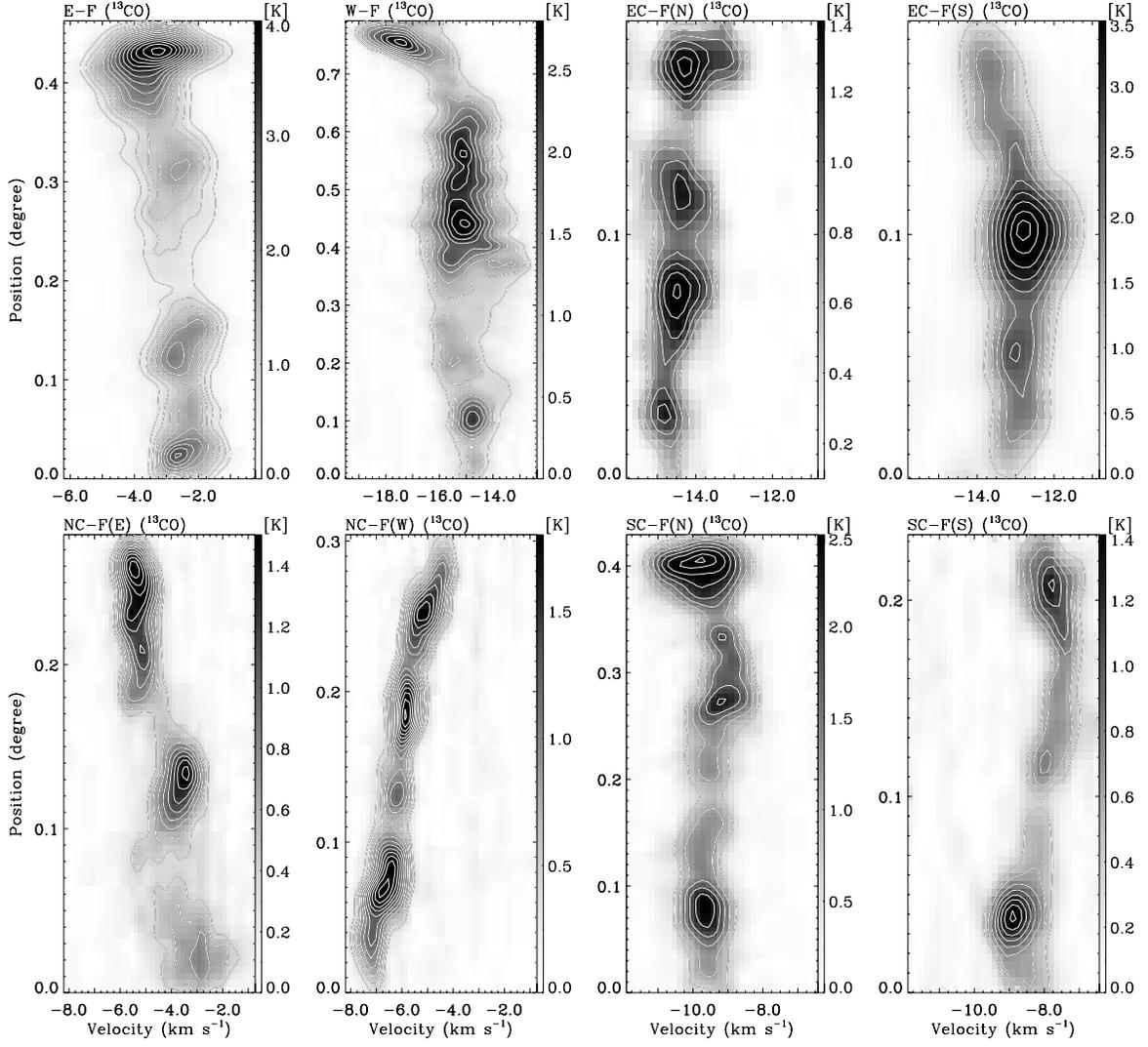}
\caption{The $^{13}$CO $PV$ plots of the identified filaments extracted along the directions of the solid arrowed lines and with the widths of the dashed boxes shown in Figures~\ref{fig:f6}, \ref{fig:f8}, \ref{fig:f10}, \ref{fig:f12}, and \ref{fig:f14}. The contours are overlaid from 10~$\sigma$ at intervals of 5~$\sigma$ for E-F, W-F, EC-F(S), and SC-F(N) and from 5~$\sigma$ at intervals of 2~$\sigma$ for the rest ($\sigma$ is the rms noise level in each plot).}
\label{fig:f15}
\end{figure}

\begin{figure}
\epsscale{1.0}
\plotone{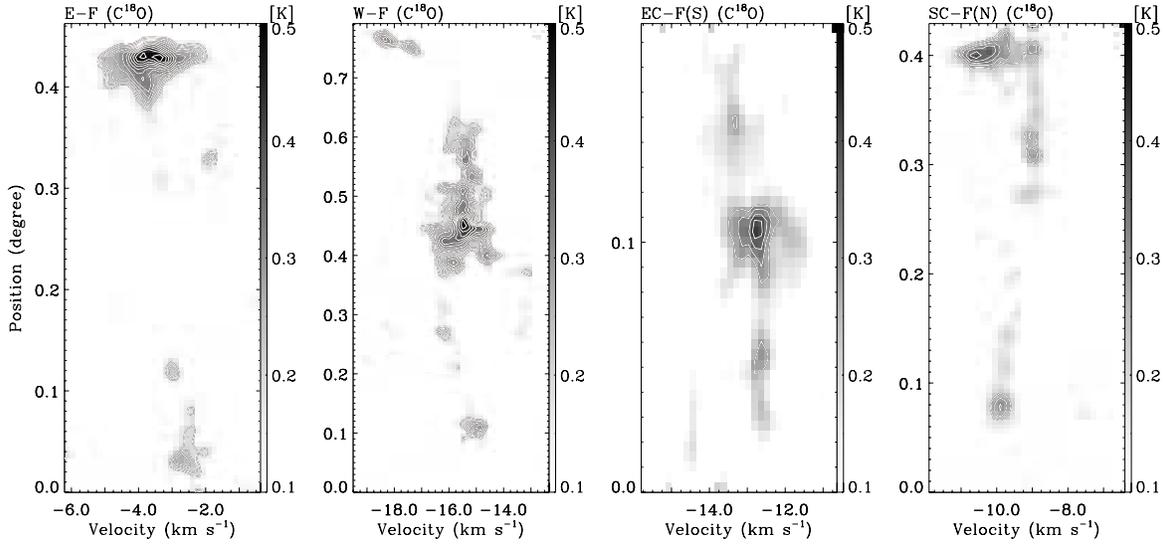}
\caption{The C$^{18}$O $PV$ plots of the identified filaments extracted along the directions of the solid arrowed lines and with the widths of the dashed boxes shown in Figure~\ref{fig:f6}, \ref{fig:f8}, \ref{fig:f12}, and \ref{fig:f14}. The contours are overlaid from 3~$\sigma$ at intervals of 0.5~$\sigma$ for E-F, W-F, EC-F(S), and SC-F(N) ($\sigma$ is the rms noise level in each plot).}
\label{fig:f16}
\end{figure}

\begin{figure}
\epsscale{1.0}
\plotone{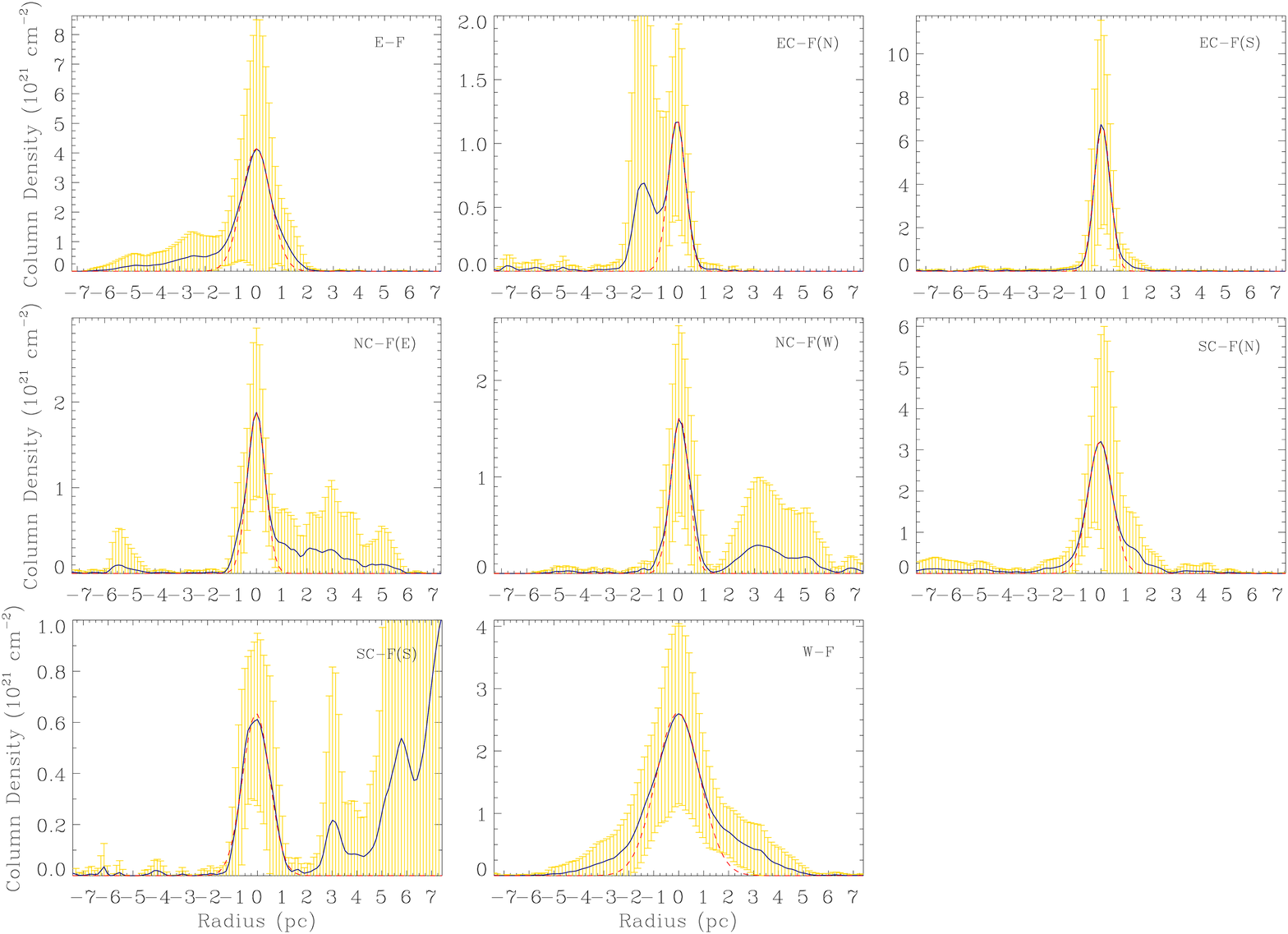}
\caption{Mean radial column density profiles perpendicular to the filamentary structures (navy curves). The position of the peak density in each profile is regarded as the center of the profile and thus the position of $R = 0$. The $R < 0$ and $R > 0$ parts of each profile correspond to the left and right sides of each filament. The yellow areas show the $\pm$~1~$\sigma$ dispersion of the distributions of radial profiles along the filaments. The dashed red curves show the Gaussian fittings to the inner parts of the profiles.}
\label{fig:f17}
\end{figure}

We further calculate the radial density profiles of filaments based on the H$_2$ column density (shown in panel (d2) of Figures~\ref{fig:f6}, \ref{fig:f8}, \ref{fig:f10}, \ref{fig:f12}, and \ref{fig:f14}). Using the same method as \citet{pal13} and \citet{xio17}, we first determine the tangential direction of each pixel along the position of each filament, shown as the solid white line in each (d2) panel. For each pixel, we then derive one column density profile perpendicular to the tangential direction. We average the profiles of all pixels along each filament and finally derive the mean radial density profile. The results are presented in Figure~\ref{fig:f17}. Most profiles have Gaussian-like shapes in the inner parts, and the outer parts of the profiles reflect the distributions of the surrounding molecular gas. We note that EC-F(N) has two Gaussian components; the one at $\sim$~$-$1.5~pc actually reflects the molecular gas belonging to EC-F(S) that is located at the same Galactic latitude of the southern end of EC-F(N) (see Figure~\ref{fig:f8}). The inner profile of SC-F(S) is much weaker than the outer profile ($R > 3$~pc), which reflects the molecular gas from the ``right'' filament SC-F(N) (see Figure~\ref{fig:f12}).

As pointed out by the works of the $Herschel$ Gould Belt survey \citep[e.g.,][]{arz11, pal13, cox16}, filamentary structures tend to have a narrow distribution in width with a median value of 0.10$\pm$0.03~pc. However, some recent works \citep[e.g.,][]{hac18, oss19} show that the width of the filament may not be universal; it could be smaller or larger than 0.1~pc. In this work, we also calculate the FWHM width of each filament by applying Gaussian fits to the inner parts of the profiles (shown as the dashed red curves in Figure~\ref{fig:f17}). The results are listed in Table\ref{tab:tab2}. The mean width from these eight filaments is 1.13$\pm$0.01~pc, which is much larger than 0.1~pc but similar to our previous work \citep[0.79~pc;][]{xio17}. We note that the resolution of our observations (50$\arcsec$, corresponding to $\sim$0.4~pc at a distance of $\sim$1.7~kpc) may not be sufficient enough to resolve the inner width of the filaments. Further high-resolution observations are needed to have a better understanding of the filament width in this region.

We also apply the Plummer fits \citep[e.g.,][]{arz11, pal13} to the radial density profiles; however, the Plummer-like function does not fit the profiles well, and the power index $p$ turns out to be much larger than 2 (the value frequently seen in the $Herschel$ Gould Belt survey). We note that the Plummer-like function tends to fit the profile in a broader range than the Gaussian function, while the outer parts of the profiles could be easily affected by the line emission from the surrounding molecular gas.

\subsubsection{Other properties of filaments}
In addition, we calculate the $^{13}$CO line width of each filament. Following the method mentioned in Section~\ref{sec:obs}, we first derive the line width of each pixel in the area around each filament where the integrated intensity of the $^{13}$CO emission is greater than three times the rms noise. Then, we use the integrated intensity of these pixels as the weight to calculate the averaged line width of each area and thus the line width of each filament. The results, along with the mean $T_{\rm ex}$ and $N_{\rm H_2}$ derived in previous sections, are listed in Table~\ref{tab:tab2}.

\section{DISCUSSION}\label{sec:disc}

\subsection{Gravitational Stability of Filaments}\label{sec:disc:stab}
According to the framework of \citet{cha53} and the following studies \citep{ost64, nag87, fie00, jac10, fis12}, an isothermal gas cylinder would go through ``sausage instability'' and fragment along the radial direction once its mass per unit length (thus, line mass) exceeds a critical value of
\begin{equation}
M_{\rm crit} = 2\sigma^{2}/G,
\end{equation}
where $\sigma$ is the velocity dispersion representing the internal pressure \citep{wan14}, and $G$ is the gravitational constant, which is given as 1/232~km$^2$~s$^{-2}$~$M_{\sun}^{-1}$~pc \citep{sol87}. Therefore, the gravitational stability of a filament is determined by this critical mass. If the line mass of the filament is greater than the critical value, the filament is gravitationally unstable and will fragment into clumps along its length that may lead to star formation \citep[e.g.,][]{gol08, bon10, xio17}. On the contrary, with the line mass lower than the critical value, the filament is gravitationally unbound or in an expanding state and is even expected to disperse during the turbulent crossing time \citep{arz13}, unless confined by the external pressure \citep{fis12}.

The key parameter in the formula of critical mass is $\sigma$. When the internal support of the filament is mainly from thermal pressure, $\sigma$ is substituted by the isothermal sound speed $c_{\rm s}$, and the critical mass becomes $M_{\rm thermal} = 2c_{\rm s}^{2}/G$ \citep{ost64}. The isothermal sound speed can be derived from
\begin{equation}
c_{\rm s} = \sqrt{\frac{k_{\rm B} T_{\rm kin}}{\mu_{\rm p} m_{\rm H}}},
\end{equation}
where $\mu_{\rm p}$ is the mean molecular weight per free particle with a value of 2.33 \citep{kau08}. In this work, $T_{\rm kin}$ is regarded as $T_{\rm ex}$ under the LTE condition (see Section~\ref{sec:res:east}). In case of turbulence dominating the internal support, $\sigma$ is replaced by the total velocity dispersion $\sigma_{\rm tot}$, which can be calculated as
\begin{equation}
\sigma_{\rm tot} = \sqrt{c_{\rm s}^{2} + \sigma_{\rm NT}^{2}} = \sqrt{c_{\rm s}^{2} + \sigma_{\rm obs}^{2} - \sigma_{\rm T,obs}^{2}}.
\end{equation}
Here $\sigma_{\rm NT}$ is the nonthermal velocity dispersion, which can be obtained by subtracting the thermal velocity dispersion ($\sigma_{\rm T,obs}$) from the observed velocity dispersion ($\sigma_{\rm obs}$). We have derived the $^{13}$CO line width ($\Delta v$) of the filaments, so the observed velocity dispersion ($\sigma_{\rm obs}$) is $\frac{\Delta v}{\sqrt{8 \rm ln 2}}$. The thermal velocity dispersion ($\sigma_{\rm T,obs}$) is $\sqrt{\frac{k_{\rm B} T_{\rm kin}}{\mu_{\rm obs} m_{\rm H}}}$. Therefore, the critical line mass becomes the virial line mass per unit length $M_{\rm Vir} = 2\sigma_{\rm tot}^{2}/G$ \citep{fie00}.

\begin{figure}
\epsscale{1.0}
\plotone{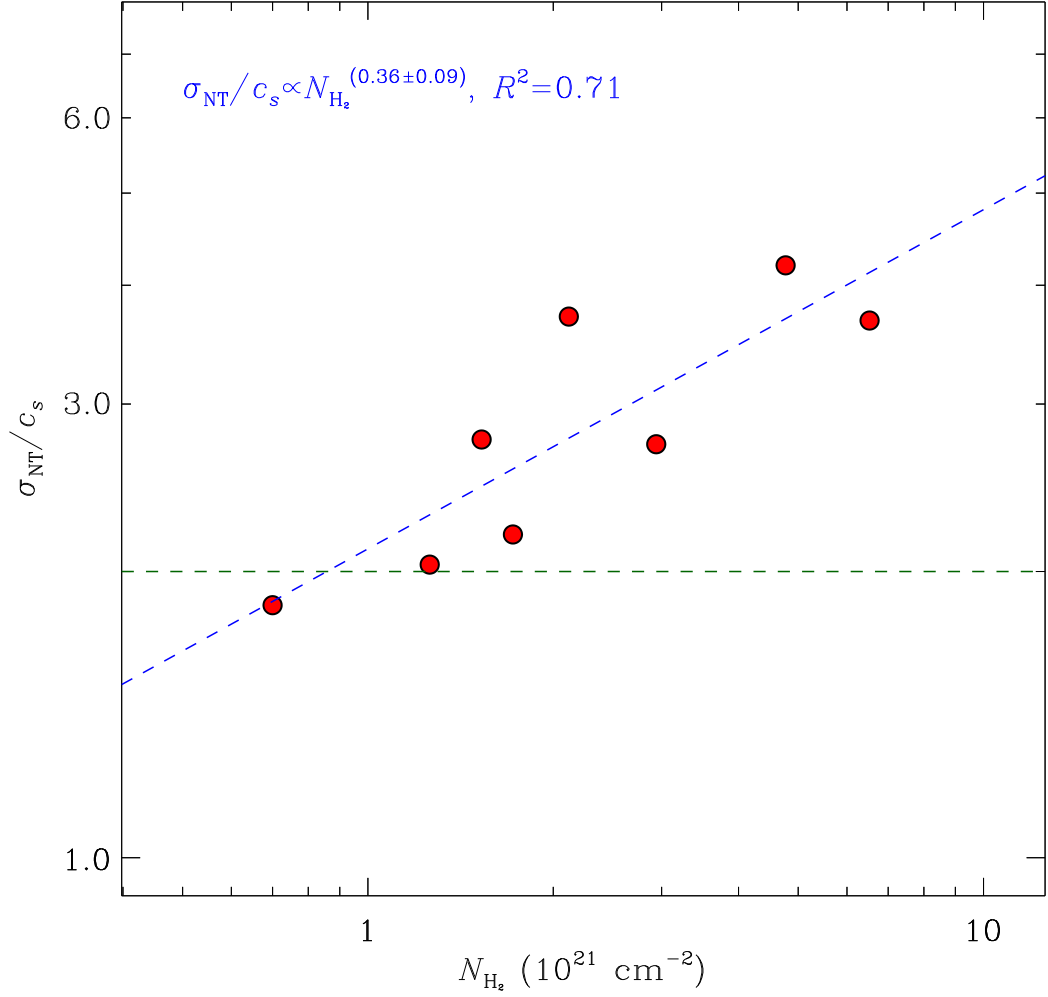}
\caption{Relationship between the ratio of nonthermal velocity dispersion ($\sigma_{\rm NT}$) to isothermal sound speed ($c_{\rm s}$) and the mean H$_2$ column density ($N_{\rm H_2}$) traced by the $^{13}$CO emission of filaments. The dashed green line marks the position of $\sigma_{\rm NT}/c_{\rm s} = 2$. The dashed blue line is the power-law fitting of the correlation. Here $R^{2}$ is the correlation coefficient in the unit of percentage.}
\label{fig:f18}
\end{figure}

By calculating these parameters, we compare our observations with scenarios of thermal or turbulent support in the filaments. The results are listed in Table~\ref{tab:tab2}. Figure~\ref{fig:f18} illustrates the relationship between the ratio of nonthermal velocity dispersion to isothermal sound speed and the mean H$_2$ column density traced by the $^{13}$CO emission in the filaments. The ratio ($\sigma_{\rm NT}/c_{\rm s}$) ranges from 1.84 to 4.20. For most filaments, the nonthermal (turbulent) motions are supersonic ($\sigma_{\rm NT}/c_{\rm s} > 2$; see the dashed green line in Figure~\ref{fig:f18}). Basically, the ratio ($\sigma_{\rm NT}/c_{\rm s}$) increases with the mean column density ($N_{\rm H_2}$), and the relationship can be fitted with a power law of $\sigma_{\rm NT}/c_{\rm s} \propto N_{\rm H_2}^{(0.36 \pm 0.09)}$. We further calculate the $M_{\rm thermal}$ of each filament; the results are 9.83 (E-F), 9.39 (EC-F(N)), 8.21 (EC-F(S)), 10.18 (NC-F(E)), 11.59 (NC-F(W)), 10.09 (SC-F(N)), 7.35 (SC-F(S)), and 11.58 (W-F) in units of $M_\sun$ pc$^{-1}$. Compared to the LTE line mass, we find that all of the filaments are thermally supercritical, with the LTE mass ($M_{\rm LTE}$) much larger than the critical mass ($M_{\rm thermal}$) by factors of 2.55--15.62. We also calculate the virial line mass ($M_{\rm Vir}$) of each filament (see the last column of Table~\ref{tab:tab2}), which is more consistent with the observed LTE line mass. Clearly, in our observations, thermal pressure is not enough to support the fragmentation of filaments. The turbulence should be the dominant support against the radial collapse. Therefore, the instability along the filaments and subsequent fragmentation can develop \citep{jac10, wan14, beu15, lux18}.

\begin{figure}
\epsscale{1.0}
\plotone{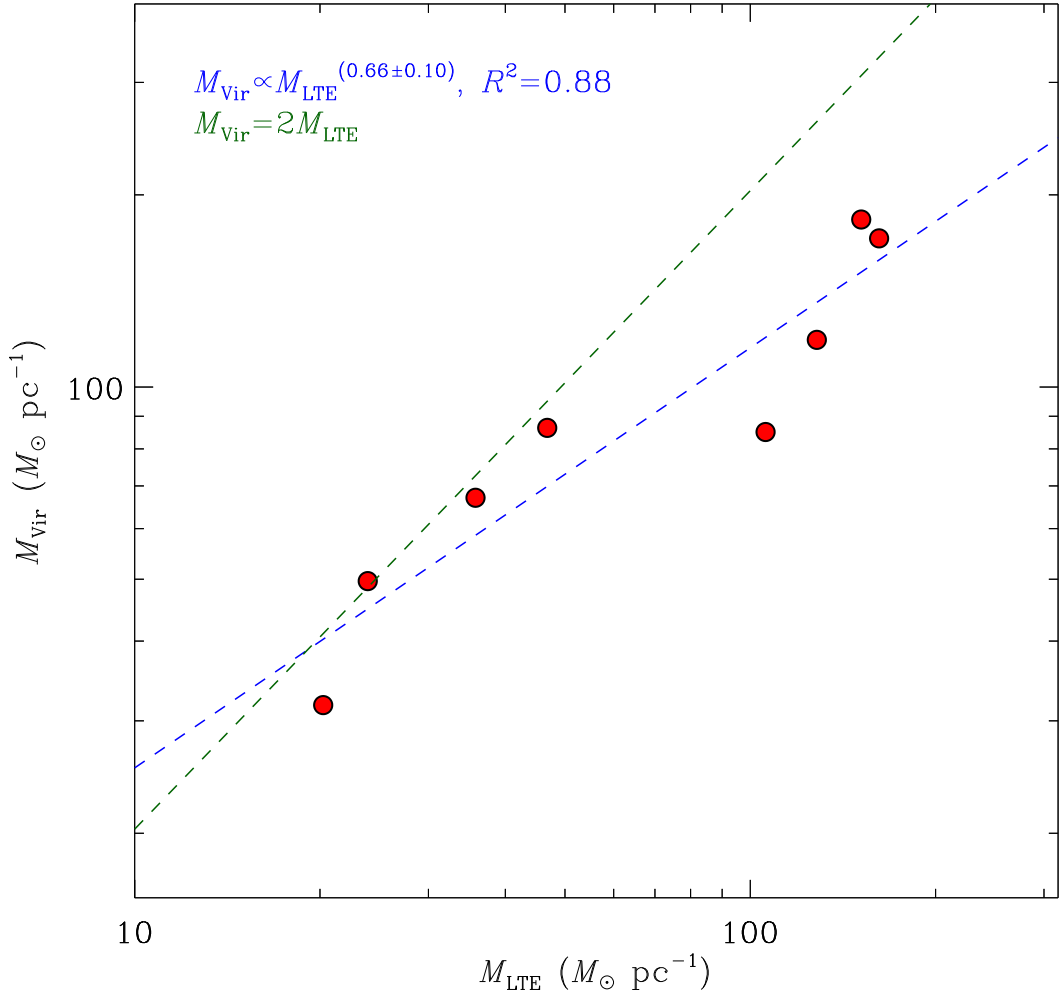}
\caption{Relationship between the virial line mass ($M_{\rm Vir}$) and the LTE line mass ($M_{\rm LTE}$) of the filaments. The dashed blue line is the power-law fitting of the correlation. The dashed green line marks the position where the virial parameter equals 2. Here $R^{2}$ is the correlation coefficient in the unit of percentage.}
\label{fig:f19}
\end{figure}

In Figure~\ref{fig:f19}, we present the relationship between virial line mass and LTE line mass. Only EC-F(N) has a virial parameter ($\alpha_{\rm Vir} = M_{\rm Vir}/M_{\rm LTE}$) larger than 2, with a value of 2.04; thus, it tends to be gravitationally unbound and unvirialized. All of the rest of the filaments have virial parameters smaller than 2, and four of them have a virial line mass close to the LTE line mass: E-F ($\alpha_{\rm Vir} = 1.21$), EC-F(S) ($\alpha_{\rm Vir} = 0.92$), SC-F(N) ($\alpha_{\rm Vir} = 0.80$), and W-F ($\alpha_{\rm Vir} = 1.06$). These four filaments are the ones with detectable C$^{18}$O emission (see Section~\ref{sec:res}), indicating that they are more condensed than the other filaments and more likely to be gravitationally bound. \citet{arz13} also found that the thermally supercritical filaments detected by the $Herschel$ Gould Belt Survey are self-gravitating structures in a rough virial balance with $M_{\rm Vir} \sim M_{\rm LTE}$. Shown in Figure~\ref{fig:f19}, the correlation can be fitted as a power law of $M_{\rm Vir} \propto M_{\rm LTE}^{0.66 \pm 0.10}$, which is also similar to the result ($M_{\rm Vir} \propto M_{\rm LTE}^{0.76}$) for the gravitationally bound filaments in \citet{arz13}.

\subsection{CO Clumps along Filaments}
To further study fragmentation within the filaments, we adopt the $Clumpfind$ algorithm \citep{wil94} to extract the dense clumps. This algorithm searches for the local peaks of emission and follows them down to the lower intensity levels. The end result is a decomposition of the data cube into a set of structural units (``clumps'') in which the emission is concentrated. We use the $^{13}$CO data cube of each region as the input cube. We set ``low'' (the minimum value where the algorithm starts contouring the data) as 3.5 times of the rms, ``inc'' (the minimum difference between the peak emission and the lowest emission of a clump) as 1 times of the rms, and other parameters as the default values. The results are listed in Table~\ref{tab:tab3}.

Along with the clump identification, we also derive the radius, excitation temperature, H$_2$ volume density, line width, nonthermal velocity dispersion, and thermal velocity dispersion. Based on these parameters, we further calculate the clump LTE mass $M_{\rm LTE,c} = \mu m_{\rm H} (4 \pi R^{3} / 3) n$, clump virial mass $M_{\rm Vir,c} = \sigma_{\rm 3D}^{2} R / G = 3(\sigma_{\rm NT}^{2}+c_{\rm s}^{2}) R / G$ \citep{mac88, wil94}, and clump Jeans mass $M_{\rm Jeans,c} = \pi^{5/2} c_{\rm s}^{3} / 6 \sqrt{G^{3} \mu m_{\rm H} n}$ \citep{gib09, wan14}. We associate these clumps, whose positions and velocities are within the emission area and velocity component of the filament, with the filament. The results are shown in the last column of Table~\ref{tab:tab3}.

\begin{figure}
\epsscale{1.0}
\plotone{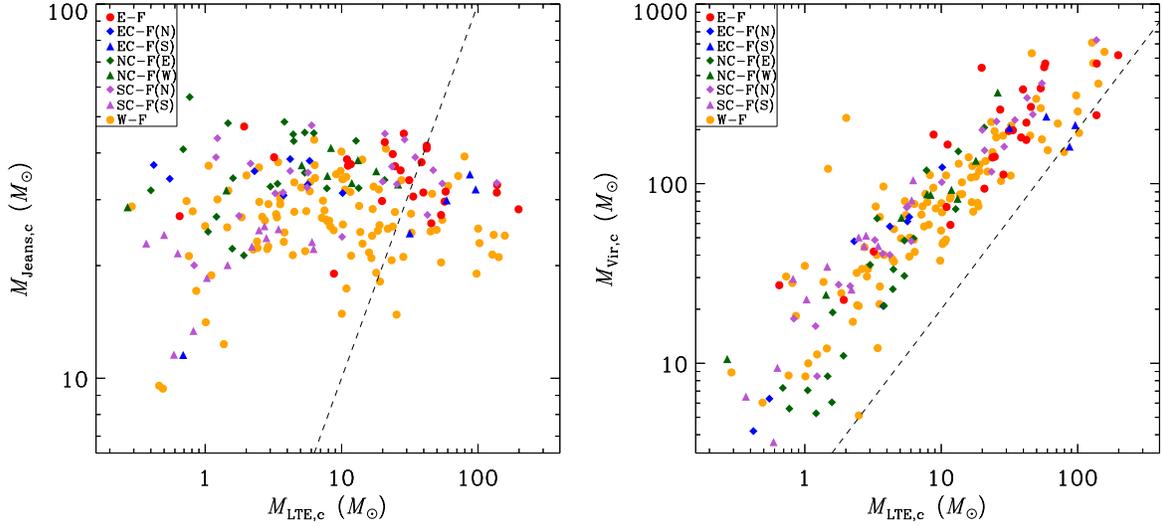}
\caption{Left panel: relationship between the Jeans mass ($M_{\rm Jeans,c}$) and the LTE mass ($M_{\rm LTE,c}$) of the clumps located on the filaments. Different shapes and colors represent the clumps located on different filaments, which are marked in the top right corner of the plot. The dashed black line indicates the position where Jeans mass equals LTE mass. Right panel: relationship between the virial mass ($M_{\rm Vir,c}$) and the LTE mass ($M_{\rm LTE,c}$) of the clumps located on the filaments. The dashed black line indicates the position where the virial parameter equals 2.}
\label{fig:f20}
\end{figure}

In Figure~\ref{fig:f20}, we present correlations between different masses ($M_{\rm Jeans,c}$, $M_{\rm Vir,c}$, and $M_{\rm LTE,c}$) of clumps located on the filaments. For most of the clumps, both the Jeans mass and the virial mass are larger than the LTE mass, indicating that the thermal and nonthermal motions (e.g., turbulence) are working together to support the clumps against the gravity. We also find that the LTE mass is more correlated with the virial mass, as they present a robust power-law relationship. However, there is no clear correlation between the LTE mass and the Jeans mass.

\begin{figure}
\epsscale{1.0}
\plotone{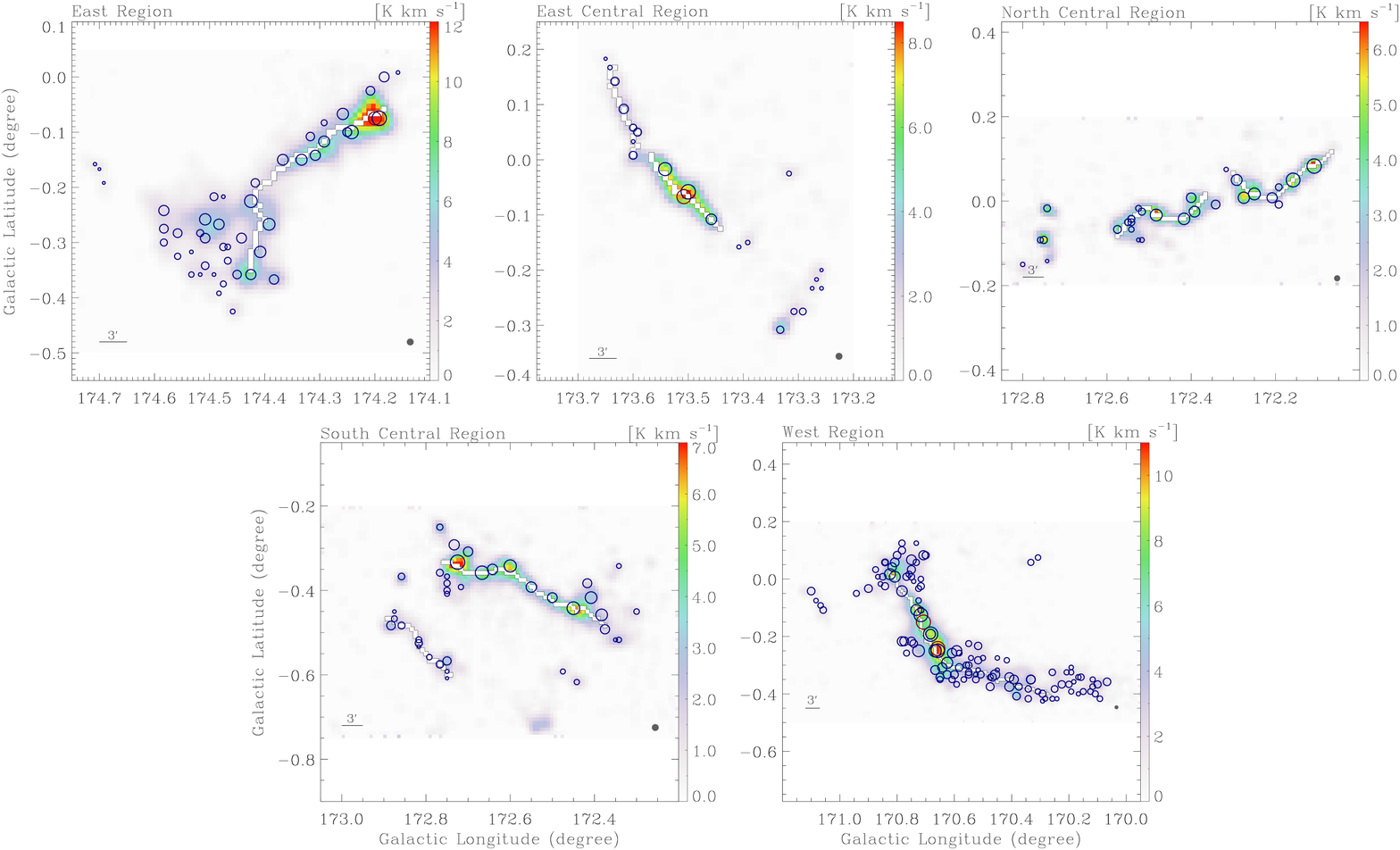}
\caption{The CO clumps identified in each region. The circles indicate the positions of the clumps. The sizes of the circles are scaled according to the sizes of the clumps. The red circles indicate the virialized clumps, while the blue circles indicate the unvirialized clumps. The solid white lines show the positions of filaments. The backgrounds are the $^{13}$CO integrated intensity maps of each region.}
\label{fig:f21}
\end{figure}

Figure~\ref{fig:f21} shows the positions of clumps identified in each region. We mark the virialized clumps ($M_{\rm Vir,c} / M_{\rm LTE,c} < 2$) with red circles and unvirialized clumps ($M_{\rm Vir,c} / M_{\rm LTE,c} > 2$) with blue circles. Previous works \citep{and10, men10} suggested that the spatial distribution of dense cores is closely related to the filamentary network. In IRDC G011.11$-$0.12 \citep{hen10} and the Aquila cloud \citep{kon15}, about 75$\%$ of the detected protostellar cores are found to be located within the filamentary structures. In our case, among the clumps identified in each region, $\sim$64$\%$ of them, on average, are located on the filamentary structures. The virialized clumps are located at the northern end of E-F, the middle part of EC-F(S), and the middle part of W-F.

We also find that the virialized clumps located on E-F and W-F are consistent with the locations of the velocity oscillations (see Figure~\ref{fig:f15}) of the filaments. As pointed out by \citet{hac11}, the velocity oscillations corresponding to the dense cores could indicate the core-forming motions during the fragmentation of the filaments. This is consistent with our result in the last section that E-F and W-F are gravitationally bound, and subsequent fragmentation could happen within the filaments.

\subsection{YSO candidates along Filaments}
\begin{figure}
\epsscale{1.0}
\plotone{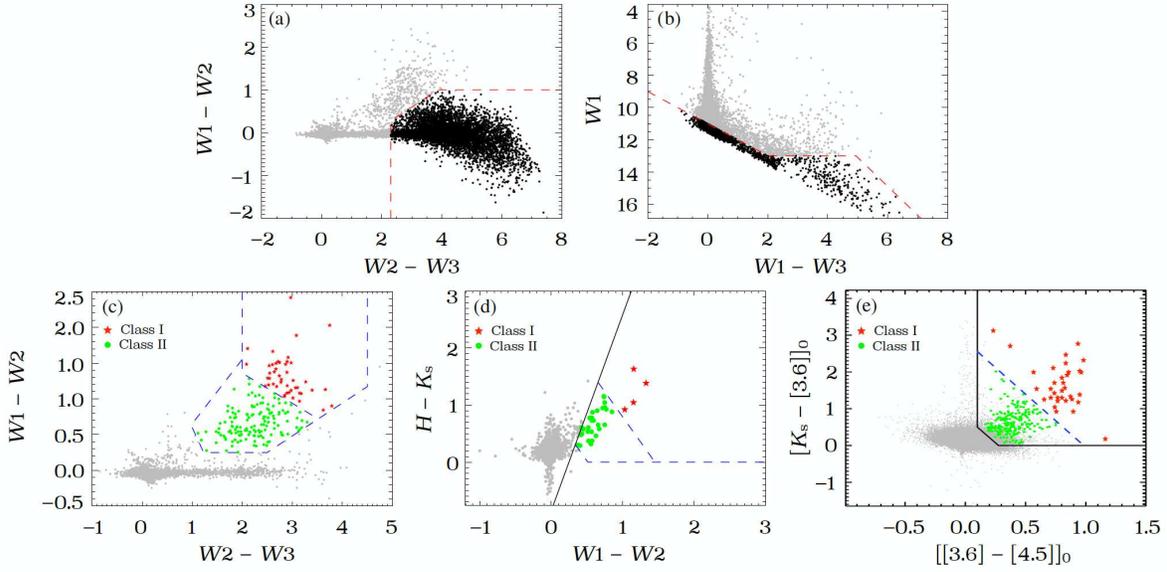}
\caption{Top panels: color-color diagrams used to remove the star-forming galaxies and AGNs. The dashed red lines indicate the criteria used in \citet{koe14}. The black points represent the sources that have been excluded, and the gray points represent the sources that have been selected for further classification. Bottom panels: classification of YSO candidates in the $W1 - W2$ vs. $W2 - W3$ color-color diagram (panel (c)), $H - K_{\rm s}$ vs. $W1 - W2$ color-color diagram (panel (d)), and $K_{\rm s} - [3.6]$ vs. $[3.6] - [4.5]$ color-color diagram (panel (e)). The solid black lines and dashed blue lines represent the classification criteria used in \citet[][panels (c) and (d)]{koe14} and \citet[][panel (e)]{gut09}. The gray points represent the sources that have been excluded.}
\label{fig:f22}
\end{figure}

With the IR data from the 2MASS, $WISE$, and GLIMPSE 360 surveys, we investigate YSO candidates in these five regions. The YSOs can be classified into disk-bearing YSOs and diskless YSOs according to the presence of circumstellar disks. The IR emission excess created by dusty circumstellar disks makes the IR colors of disk-bearing YSOs different from those of diskless YSOs. The diskless YSOs are unable to be identified only based on their IR colors. So in this work, we only focus on the disk-bearing YSOs, namely Class~I and Class~II objects. Following the scheme provided by \citet{koe14}, we first remove the star-forming galaxies and broad-line active galactic nuclei (AGNs) as extragalactic contaminants according to their locations in the $WISE$ $W1 - W2$ versus $W2 - W3$ and $W1$ versus $W1 - W3$ color-color diagrams (top panels of Figure~\ref{fig:f22}). Then we select the YSO candidates based on their locations in the $WISE$ $W1 - W2$ versus $W2 - W3$ color-color diagram, which is shown in panel (c) of Figure~\ref{fig:f22}. With the 2MASS $H$ and $K_{\rm s}$ bands, we use the $H - K_{\rm s}$ versus $W1 - W2$ color-color diagram (panel (d) of Figure~\ref{fig:f22}) to search for the YSO candidates among previously unclassified objects. $WISE$ $W4$ photometry is also introduced to identify transition disks and retrieve protostars from the AGNs. Finally, we use a combination of the $WISE$ $W1 - W2$ versus $W3 - W4$ and $W1$ versus $W1 - W2$ diagrams to remove the asymptotic giant branch (AGB) stars.

Since the latitudes of our regions are within $-1\arcdeg$ to $1\arcdeg$, we further use the GLIMPSE 360 data to identify additional YSOs. Following the Phase~II classification scheme of \citet{gut09}, we first select the additional Class~I and Class~II objects according to the $K_{\rm s} - [3.6]$ versus $[3.6] - [4.5]$ color-color diagram (panel (e) of Figure~\ref{fig:f22}). Then we adopt the criteria in \citet{sar15} to remove the extragalactic contaminants and AGB stars. We combine these additional YSOs with the YSOs identified from $WISE$ data and list the IR photometric magnitudes and classifications of all of the YSOs in Table~\ref{tab:tab4}.

\begin{figure}
\epsscale{1.0}
\plotone{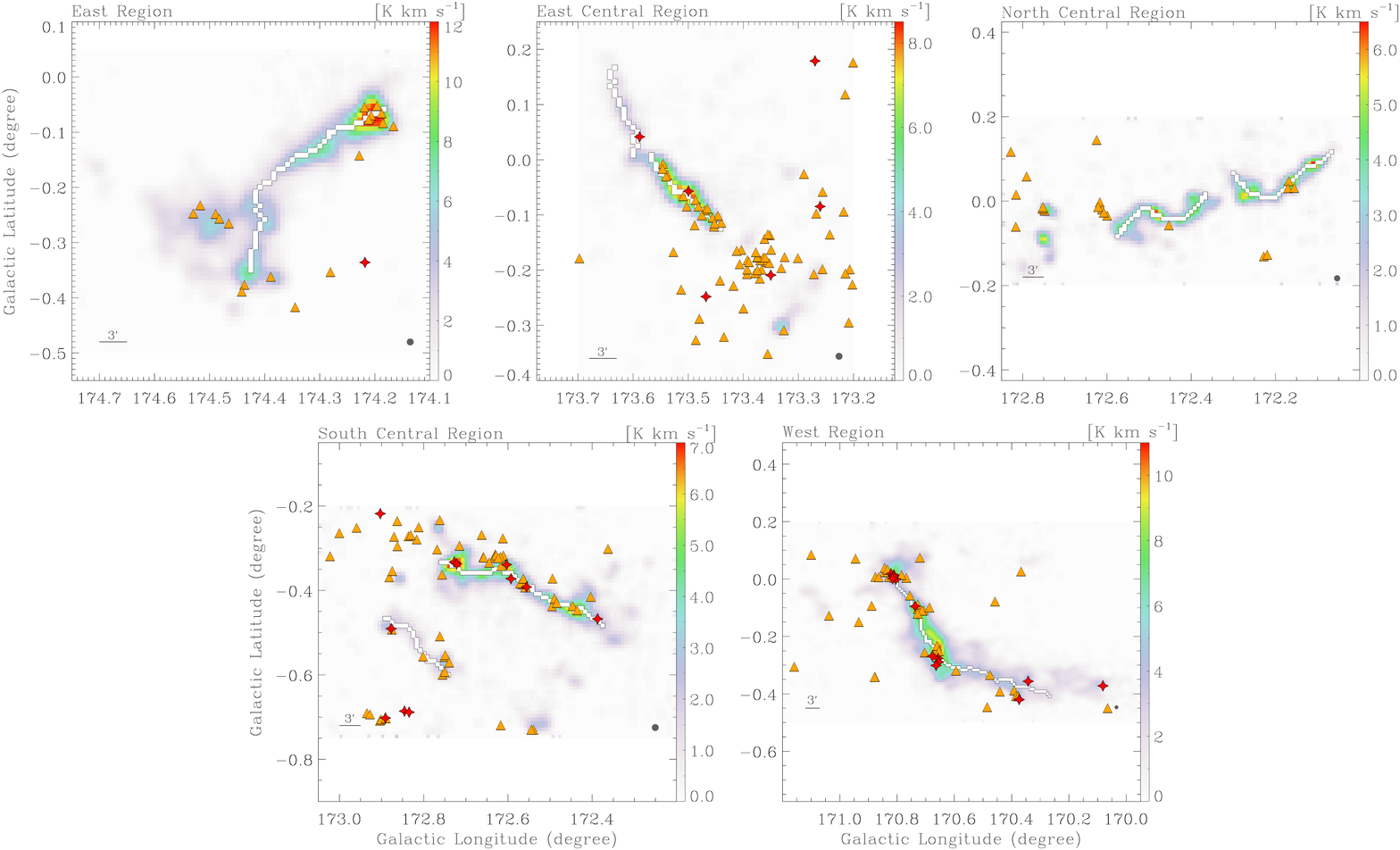}
\caption{Distributions of YSO candidates identified in each region. The red stars represent Class~I objects, and the yellow triangles represent Class~II objects. The solid white lines show the positions of filaments. The backgrounds are the $^{13}$CO integrated intensity maps of each region.}
\label{fig:f23}
\end{figure}

As shown in Figure~\ref{fig:f23}, there are two groups of Class~II objects located at the two ends of E-F. No Class~I object is identified to be associated with E-F. In the eastern central region, one Class~I object is found to be located on EC-F(N), and another one is located at the center of EC-F(S), where the virialized clump is also identified (see Figure~\ref{fig:f21}). A number of Class~II objects are also located along EC-F(S). In the northern central region, only a few Class~II objects are found near the central parts of NC-F(E) and NC-F(W). For SC-F(N), about six Class~I objects are located near the center and two ends. At one end of SC-F(S) are located one Class~I and one Class~II object, and at the other end is a group of Class~II objects. For W-F, a number of Class~I objects are identified near four locations ($l \sim 170\arcdeg.81, b \sim 0\arcdeg.01$), ($l \sim 170\arcdeg.74, b \sim -0\arcdeg.09$), ($l \sim 170\arcdeg.65, b \sim -0\arcdeg.28$), and ($l \sim 170\arcdeg.35, b \sim -0\arcdeg.38$). At each location, a group of Class~II objects is located around the Class~I objects.

We also note that the distributions of YSOs are related to the emission areas with higher temperatures ($T_{\rm ex}$ $>$ 15~K) of the filaments (e.g., E-F and W-F). Moreover, the filaments associated with more YSOs tend to have higher temperatures than the ones with fewer YSOs (e.g., EC-F(S) and EC-F(N), SC-F(N) and SC-F(S)). As pointed out by \citet{gon16}, the star-forming molecular gas tends to have a higher excitation temperature than the non-star-forming molecular gas. In our case, the existing YSOs could be the possible internal heating sources of the molecular gas with higher temperature.

\subsection{Outflows within Filaments}
\begin{figure}
\epsscale{1.0}
\plotone{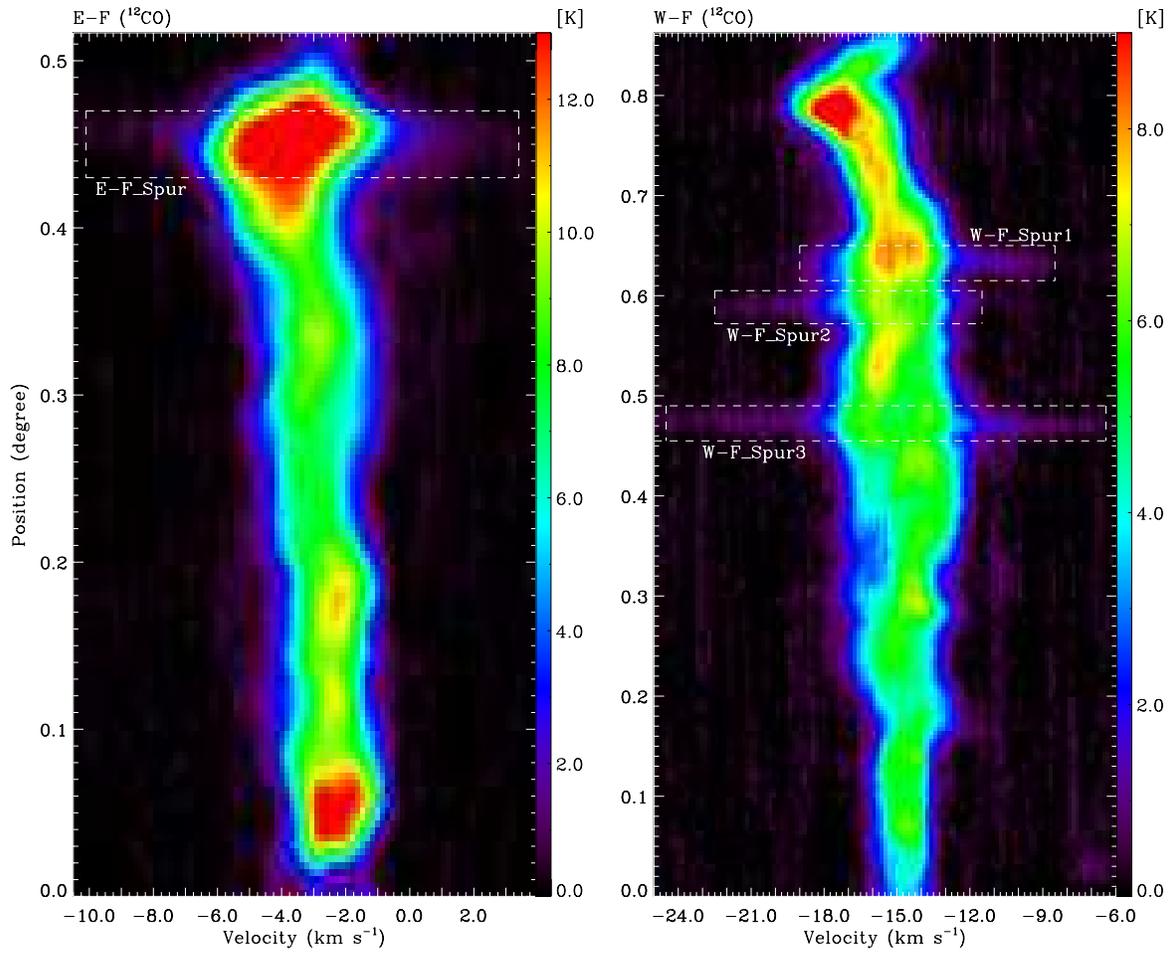}
\caption{The $^{12}$CO $PV$ plots of E-F and W-F extracted along the directions of the solid arrowed lines and with the widths of the dashed boxes shown in Figures~\ref{fig:f6} and \ref{fig:f14}. The dashed rectangles indicate the positions of $PV$ spur structures.}
\label{fig:f24}
\end{figure}

Using the $^{12}$CO $PV$ maps, we further search for the protostellar outflows within each filament. As shown in Figure~\ref{fig:f24}, only two filaments (E-F and W-F) are identified with evident spur structures in $PV$ diagrams. We name these structures ``E-F\underline{~~}Spur'', ``W-F\underline{~~}Spur1'', ``W-F\underline{~~}Spur2'', and ``W-F\underline{~~}Spur3'', and corresponding outflows as ``E-F\underline{~~}OF'', ``W-F\underline{~~}OF1'', ``W-F\underline{~~}OF2'', and ``W-F\underline{~~}OF3''. The velocities of the $^{13}$CO peak emission at the positions of these spur structures are regarded as the velocities of the line centers ($v_{\rm center}$). The velocity ranges of the outflow wings are derived from $^{12}$CO velocity components near the line centers where the $^{12}$CO emission intensities are greater than three times the rms noise and the corresponding $^{13}$CO emission intensities are smaller than three times the rms noise. Note that we also manually adjust the ranges of line wings based on the $^{12}$CO $PV$ maps.

\begin{figure}
\epsscale{1.0}
\plotone{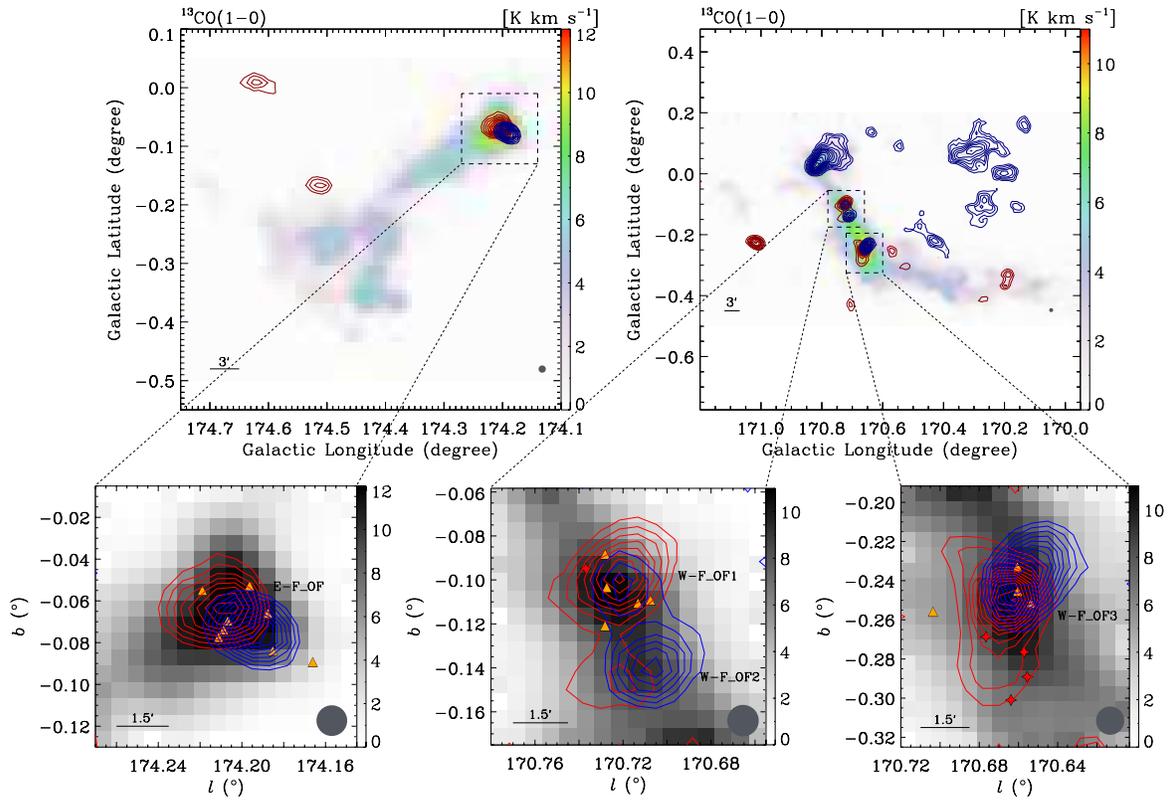}
\caption{The $^{12}$CO contours of the identified outflows with the backgrounds of the $^{13}$CO integrated intensity maps. The red contours represent the redshifted lobes of the outflows, and the blue contours represent the blueshifted lobes of the outflows. The red stars in the bottom panels represent Class~I objects, and the yellow triangles represent Class~II objects.}
\label{fig:f25}
\end{figure}

In Figure~\ref{fig:f25}, we present the CO outflows overlaid on the $^{13}$CO integrated intensity maps. The outflow images are obtained by integrating the $^{12}$CO emission over the velocity ranges of the line wings, and the contours are overlaid from 3~$\sigma$ at intervals of 1.5~$\sigma$ ($\sigma$ is the rms noise level of the outflow images). Under LTE conditions and assuming optically thin $^{12}$CO emission in the line wings, the $^{12}$CO column density of the outflow lobe can be given by \citep{sco86, gar91}
\begin{equation}
N_{\rm ^{12}CO} = 2.40 \times 10^{14} \cdot \frac{\int T_{\rm MB, ^{12}CO} dv}{1-e^{-5.53/T_{\rm ex}}},
\end{equation}
where $\int T_{\rm MB, ^{12}CO} dv$ is the integrated intensity of the outflow lobe. To derive the H$_2$ column density ($N_{\rm H_2,blue/red}$), we multiply the column density by the abundance $[{\rm H_2}]/[{\rm ^{12}CO}]$ with a value of 10$^{4}$ \citep{bla87}. The mass of the outflow lobe can be calculated by Equation~(\ref{equa:mass}),
\begin{equation}
M_{\rm bule/red} = \mu m_{\rm H} \int N_{\rm H_2,blue/red} dS_{\rm blue/red},
\end{equation}
where $S_{\rm blue/red}$ is the emission area of the blue or red lobe. The momentum and energy of the outflow are then given by 
\begin{equation}
P_{\rm out} = P_{\rm blue} + P_{\rm red} = M_{\rm bule} v_{\rm bule} + M_{\rm red} v_{\rm red},
\end{equation}
\begin{equation}
E_{\rm out} = E_{\rm blue} + E_{\rm red} = \frac{1}{2} M_{\rm bule} v^2_{\rm bule} + \frac{1}{2} M_{\rm red} v^2_{\rm red},
\end{equation}
where $v_{\rm blue/red}$ is the maximum velocity of the blue or red line wing. The results of these parameters are listed in Table~\ref{tab:tab5}. The typical mass, momentum, and energy of the outflows are 1.42~$M_\sun$, 18.72~$M_\sun$~km~s$^{-1}$, and 1.65$\times$10$^{45}$~erg, respectively. Compared to the work of \citet{lif16}, which also uses the CO data from the MWISP project to identify molecular outflows within the Galactic plane, our sample of outflows has lower mass and energy. However, as suggested by \citet{ric20} and \citet{arc07}, low-mass outflows also play an important role in the dissipation of excess angular momentum and mass accretion toward the central star during the star-forming process. 

In the bottom panels of Figure~\ref{fig:f25}, we overlay YSOs identified in the last section on the maps of outflow contours. For E-F\underline{~~}OF, W-F\underline{~~}OF1, and W-F\underline{~~}OF3, there are several Class~II objects located between the peaks of the red and blue lobes. We note that the IR data used in this work may not be able to reveal deeply embedded YSOs such as Class~0 objects, and Class~II objects are typically too evolved to power the molecular outflows in CO \citep{bac99, arc07}. Further high-resolution observations of the dust continuum are needed to identify the driving sources of these outflows. However, the detection of CO outflows indicates ongoing star-forming activity within E-F and W-F.

\section{SUMMARY}\label{sec:sum}
We present large-field mapping observations of the G172 region covering an area with $169\arcdeg.75 \leqslant l \leqslant 174\arcdeg.75$ and $-0\arcdeg.75 \leqslant b \leqslant 0\arcdeg.5$ in the $J = 1-0$ transition of the CO isotopologues ($^{12}$CO, $^{13}$CO, and C$^{18}$O) using the PMO 13.7 m telescope. We identify five main subregions of molecular clouds in the velocity range between $-$25 and 8~km~s$^{-1}$ that are characterized by conspicuous filamentary structures. The main results are summarized below.

According to the $^{13}$CO emission data of the five subregions, we have identified a total of eight filaments: E-F, EC-F(N), EC-F(S), NC-F(E), NC-F(W), SC-F(N), SC-F(S), and W-F. Four of these filaments (E-F, EC-F(S), SC-F(N), and W-F) are also revealed in the C$^{18}$O emission. We extract velocity structures along the length of each filament. In the $^{13}$CO $PV$ maps, most filaments present continuous velocity structures with slight velocity gradients. In the C$^{18}$O $PV$ maps, the velocity structures display discontinuous features, only concentrating on the densest parts.

Based on the excitation temperature (with mean values ranging from 6.16 to 9.60~K) and H$_2$ column density traced by the $^{13}$CO emission (with mean values ranging from 0.70$\times$10$^{21}$ to 6.53$\times$10$^{21}$~cm$^{-2}$) of the identified filaments, we derive the $^{13}$CO LTE line mass to be from 20.24 to 161.91~$M_\sun$ pc$^{-1}$. We obtain the radial density profiles of filaments and find that most profiles present Gaussian-like shapes in the inner parts. The Gaussian fits to the inner parts lead to a mean FWHM width of 1.13$\pm$0.01~pc.

After comparing LTE line mass with critical line mass, we find that the thermal pressure is not enough to support the fragmentation of filaments, and turbulence should be the dominant support against radial collapse. Most filaments have virial parameters smaller than 2 and thus are gravitationally bound. Four filaments (E-F, EC-F(S), SC-F(N), and W-F) have LTE line masses close to the virial line mass. The relationship of LTE mass and virial mass can be fitted as a power law of $M_{\rm Vir} \propto M_{\rm LTE}^{0.66 \pm 0.10}$.

We investigate the CO clumps in each region and find that 64$\%$ of them are associated with the filaments. The virialized clumps found in E-F and W-F are consistent with the locations of the velocity oscillations, indicating that fragmentation could happen within these gravitationally bound filaments. Based on the complementary IR data, we find that most filaments have associated Class~II objects. Only a few Class~I objects are found to be located on EC-F(S), SC-F(N), and W-F. The locations of Class~I object are also the places where virialized clumps are found.

We suggest that these four filaments, E-F, EC-F(S), SC-F(N), and W-F, may contain star-forming activity, based on the evidence of detection in the C$^{18}$O emission, virial parameters close to 1, and associated virialized clumps and Class~I objects. Especially for E-F and W-F, where we have detected $^{12}$CO outflows with typical mass, momentum, and energy of 1.42~$M_\sun$, 18.72~$M_\sun$~km~s$^{-1}$, and 1.65$\times$10$^{45}$~erg, respectively, the star-forming activity should have already taken place.

\acknowledgments
We are grateful to all of the members of the Milky Way Imaging Scroll Painting CO line survey group, especially the staff of Qinghai Radio Station of PMO at Delingha, for the support during the observations. This work was supported by the National Key Research \& Development of China under grant 2017YFA0402702 and funding from the European Union’s Horizon 2020 research and innovation program under grant agreement No. 639459 (PROMISE). We acknowledge the support from NSFC through grant Nos. 11473069, 11503086, and 11629302. This research has made use of the NASA/IPAC Infrared Science Archive, which is operated by the Jet Propulsion Laboratory, California Institute of Technology, under contract with the National Aeronautics and Space Administration.

\software{GILDAS \citep{gui00, pet05, gil13}}


\clearpage

\begin{deluxetable}{ccccccccc}
\tabletypesize{\scriptsize}
\tablewidth{0pt}
\tablecaption{Distances to Filamentary Structures in Each Region \label{tab:tab1}}
\tablehead{
\colhead{Region} & \colhead{Filament} & \colhead{$(l, b, v)_1$}           & \colhead{$d_1$} & \colhead{$(l, b, v)_2$}           & \colhead{$d_2$} & \colhead{$(l, b, v)_3$}           & \colhead{$d_3$} & \colhead{Distance} \\
                 &                    & \colhead{(deg, deg, km~s$^{-1}$)} & \colhead{(kpc)} & \colhead{(deg, deg, km~s$^{-1}$)} & \colhead{(kpc)} & \colhead{(deg, deg, km~s$^{-1}$)} & \colhead{(kpc)} & \colhead{(kpc)} \\
\colhead{(1)}    & \colhead{(2)}      & \colhead{(3)}                     & \colhead{(4)}   & \colhead{(5)}                     & \colhead{(6)}   & \colhead{(7)}                     & \colhead{(8)}   & \colhead{(9)}
}
\startdata
Eastern region                           & E-F     & (174.43, -0.36, -2.3)  & 1.68$\pm$0.12 & (174.34, -0.15, -3.0)  & 1.68$\pm$0.12 & (174.20, -0.07, -4.0)  & 1.68$\pm$0.12 & 1.68 \\
\multirow{2}{*}{Eastern central region}  & EC-F(N) & (173.63, 0.14, -15.0)  & 1.72$\pm$0.15 & (173.61, 0.09, -14.3)  & 1.72$\pm$0.15 & (173.59, 0.01, -13.7)  & 1.72$\pm$0.15 & \multirow{2}{*}{1.72} \\
                                         & EC-F(S) & (173.56, -0.01, -13.2) & 1.72$\pm$0.15 & (173.51, -0.06, -12.5) & 1.71$\pm$0.15 & (173.46, -0.11, -13.8) & 1.73$\pm$0.16 & \\
\multirow{2}{*}{Northern central region} & NC-F(E) & (172.57, -0.07, -3.0)  & 1.69$\pm$0.13 & (172.48, -0.03, -3.8)  & 1.69$\pm$0.13 & (172.38, 0.01, -5.5)   & 1.70$\pm$0.14 & \multirow{2}{*}{1.70} \\
                                         & NC-F(W) & (172.27, 0.02, -6.6)   & 1.72$\pm$0.15 & (172.16, 0.05, -5.5)   & 1.71$\pm$0.15 & (172.11, 0.09, -4.9)   & 1.71$\pm$0.14 & \\
\multirow{2}{*}{Southern central region} & SC-F(N) & (172.44, -0.44, -9.6)  & 1.76$\pm$0.20 & (172.59, -0.35, -9.2)  & 1.74$\pm$0.18 & (172.73, -0.33, -10.1) & 1.74$\pm$0.18 & \multirow{2}{*}{1.73} \\
                                         & SC-F(S) & (172.75, -0.57, -8.8)  & 1.73$\pm$0.17 & (172.82, -0.53, -8.0)  & 1.72$\pm$0.16 & (172.89, -0.48, -7.8)  & 1.72$\pm$0.15 & \\
Western region                           & W-F     & (170.38, -0.39, -14.7) & 1.90$\pm$0.24 & (170.65, -0.24, -15.0) & 1.90$\pm$0.24 & (170.82, 0.02, -17.8)  & 1.91$\pm$0.24 & 1.90 \\
\enddata
\tablecomments{Distances to the filamentary structures estimated by the Bayesian approach. Columns 3, 5, and 7 are the coordinates of the three chosen points located in each filament mentioned in Section~\ref{sec:res:pro}. Columns 4, 6, and 8 are the distances to these three points derived by the Bayesian distance estimation program. Column 9 is the distance to the filament in each region.}
\end{deluxetable}

\begin{deluxetable}{cccccccccccc}
\tabletypesize{\scriptsize}
\tablewidth{0pt}
\rotate
\tablecaption{Properties of Filamentary Structures \label{tab:tab2}}
\tablehead{
\colhead{Filament} & \colhead{Length} & \colhead{Width} & \colhead{$T_{\rm ex}$} & \colhead{$N_{\rm H_2}$ ($^{13}$CO)} & \colhead{$N_{\rm H_2}$ (C$^{18}$O)} & \colhead{$M_{\rm LTE}$ ($^{13}$CO)} & \colhead{$\Delta v$ ($^{13}$CO)} & \colhead{$\sigma_{\rm NT}$} & \colhead{$c_{\rm s}$}   & \colhead{$\sigma_{\rm tot}$} & \colhead{$M_{\rm Vir}$} \\
                   & \colhead{(pc)}   & \colhead{(pc)}  & \colhead{(K)}          & \colhead{(10$^{21}$~cm$^{-2}$)}     & \colhead{(10$^{21}$~cm$^{-2}$)}     & \colhead{($M_\sun$~pc$^{-1}$)}      & \colhead{(km~s$^{-1}$)}          & \colhead{(km~s$^{-1}$)}     & \colhead{(km~s$^{-1}$)} & \colhead{(km~s$^{-1}$)}      & \colhead{($M_\sun$~pc$^{-1}$)} \\
\colhead{(1)}      & \colhead{(2)}    & \colhead{(3)}   & \colhead{(4)}          & \colhead{(5)}                       & \colhead{(6)}                       & \colhead{(7)}                       & \colhead{(8)}                    & \colhead{(9)}               & \colhead{(10)}          & \colhead{(11)}               & \colhead{(12)}
}
\startdata
E-F      & 15.15  & 1.33$\pm$0.02   & 8.72$\pm$0.97  & 4.77$\pm$0.49  & 8.37$\pm$0.11  & 151.43  & 1.44$\pm$0.08  & 0.611  & 0.146  & 0.627  & 182.96 \\
EC-F(N)  &  7.01  & 0.82$\pm$0.01   & 8.02$\pm$0.82  & 1.26$\pm$0.16  &     $-$        &  23.91  & 0.68$\pm$0.05  & 0.289  & 0.142  & 0.323  &  48.65 \\
EC-F(S)  &  6.38  & 0.76$\pm$0.01   & 9.60$\pm$1.23  & 6.53$\pm$0.69  & 7.94$\pm$0.12  & 128.26  & 1.15$\pm$0.06  & 0.488  & 0.133  & 0.505  & 118.54 \\
NC-F(E)  &  9.52  & 0.83$\pm$0.01   & 8.93$\pm$0.96  & 1.53$\pm$0.19  &     $-$        &  46.79  & 0.95$\pm$0.06  & 0.407  & 0.148  & 0.431  &  86.29 \\
NC-F(W)  & 11.01  & 0.84$\pm$0.01   & 9.41$\pm$1.01  & 1.72$\pm$0.23  &     $-$        &  35.78  & 0.82$\pm$0.05  & 0.345  & 0.158  & 0.380  &  67.07 \\
SC-F(N)  & 14.59  & 1.13$\pm$0.01   & 8.99$\pm$0.99  & 2.94$\pm$0.32  & 5.05$\pm$0.07  & 105.88  & 0.94$\pm$0.05  & 0.401  & 0.148  & 0.428  &  85.02 \\
SC-F(S)  &  8.05  & 1.29$\pm$0.01   & 6.16$\pm$0.49  & 0.70$\pm$0.09  &     $-$        &  20.24  & 0.54$\pm$0.04  & 0.231  & 0.261  & 0.270  &  31.75 \\
W-F      & 28.45  & 2.07$\pm$0.03   & 8.13$\pm$0.74  & 2.12$\pm$0.25  & 3.52$\pm$0.05  & 161.91  & 1.38$\pm$0.08  & 0.585  & 0.158  & 0.606  & 170.93 \\
\enddata
\tablecomments{Properties of the identified filaments, including the length of each filament (column 2), width of each filament (column 3), excitation temperature (column 4) and line width in the $^{13}$CO emission (column 8). Columns 5--6 are mean H$_2$ column density traced by $^{13}$CO and C$^{18}$O. Columns 9--11 are the nonthermal, thermal, and total velocity dispersion. Columns 7 and 12 are the LTE and virial line mass, respectively.}
\end{deluxetable}

\begin{longrotatetable}
\begin{deluxetable}{cccccccccccccc}
\tabletypesize{\tiny}
\tablewidth{0pt}
\tablecaption{Properties of Clumps in Each Region \label{tab:tab3}}
\tablehead{
\colhead{Clump} & \colhead{Velocity}      & \colhead{Radius} & \colhead{$T_{\rm ex}$} & \colhead{$n_{\rm H_2}$}        & \colhead{$M_{\rm LTE,c}$} & \colhead{$\Delta v$}    & \colhead{$\sigma_{\rm NT}$} & \colhead{$c_{\rm s}$}   & \colhead{$\sigma_{\rm 3D}$} & \colhead{$M_{\rm Vir,c}$} & \colhead{$M_{\rm Jeans,c}$} & \colhead{Region} & \colhead{Filament} \\
                & \colhead{(km~s$^{-1}$)} & \colhead{(pc)}   & \colhead{(K)}          & \colhead{(10$^{2}$~cm$^{-3}$)} & \colhead{($M_\sun$)}      & \colhead{(km~s$^{-1}$)} & \colhead{(km~s$^{-1}$)}     & \colhead{(km~s$^{-1}$)} & \colhead{(km~s$^{-1}$)}     & \colhead{($M_\sun$)}      & \colhead{($M_\sun$)} \\
\colhead{(1)}   & \colhead{(2)}           & \colhead{(3)}    & \colhead{(4)}          & \colhead{(5)}                  & \colhead{(6)}             & \colhead{(7)}           & \colhead{(8)}               & \colhead{(9)}           & \colhead{(10)}              & \colhead{(11)}            & \colhead{(12)}              & \colhead{(13)}   & \colhead{(14)}
}
\startdata
MWISP G174.200$-$00.075 &  -4.21 & 1.339 & 14.36 & 2.32 & 137.35 & 0.927 & 0.388 & 0.226 & 0.778 & 240.52 & 31.34 & Eastern & E-F \\
MWISP G174.192$-$00.075 &  -3.71 & 1.499 & 12.82 & 2.03 & 198.21 & 1.287 & 0.543 & 0.213 & 1.011 & 519.12 & 28.28 & Eastern & E-F \\
MWISP G174.383$-$00.367 &  -3.05 & 0.978 & 13.28 & 1.14 &  23.70 & 0.827 & 0.346 & 0.217 & 0.707 & 139.87 & 39.72 & Eastern & E-F \\
MWISP G174.425$-$00.358 &  -2.05 & 1.069 & 15.62 & 1.68 &  42.01 & 0.885 & 0.370 & 0.235 & 0.759 & 175.11 & 41.80 & Eastern & E-F \\
MWISP G174.425$-$00.225 &  -1.88 & 1.324 & 11.86 & 0.76 &  42.06 & 0.888 & 0.373 & 0.205 & 0.737 & 218.63 & 41.11 & Eastern & E-F \\
MWISP G174.392$-$00.267 &  -3.05 & 1.234 & 12.09 & 0.95 &  38.52 & 0.837 & 0.350 & 0.207 & 0.705 & 180.84 & 37.78 & Eastern & E-F \\
MWISP G174.308$-$00.142 &  -3.05 & 1.061 & 11.88 & 1.92 &  45.45 & 1.098 & 0.462 & 0.205 & 0.876 & 267.47 & 25.94 & Eastern & E-F \\
MWISP G174.450$-$00.358 &  -3.05 & 0.884 & 15.51 & 1.57 &  20.72 & 0.712 & 0.295 & 0.235 & 0.653 &  93.80 & 42.75 & Eastern & E-F \\
\enddata
\tablecomments{Properties of identified clumps. Column 1 is the clump name, which contains the Galactic longitude and latitude. Columns 2 to 6 are the rest velocity, clump radius, excitation temperature, H$_2$ volume density, and LTE mass. Columns 7 to 12 are the line width, non-thermal velocity dispersion, isothermal sound speed, three-dimensional velocity dispersion, virial mass and Jeans mass. Columns 14 and 15 are the region and filament where the clump is located at.}
\end{deluxetable}
\end{longrotatetable}

\begin{longrotatetable} 
\begin{deluxetable}{cccccccccccc}
\tabletypesize{\tiny}
\tablewidth{0pt}
\tablecaption{IR Photometric Magnitudes of YSO Candidates \label{tab:tab4}}
\tablehead{
\colhead{ID} & \colhead{$J$(1.25~$\mu$m)} & \colhead{$H$(1.65~$\mu$m)} & \colhead{$K_{\rm s}$(2.17~$\mu$m)} & \colhead{$W1$(3.4~$\mu$m)} & \colhead{$W2$(4.6~$\mu$m)} & \colhead{$W3$(12~$\mu$m)} & \colhead{$W4$(22~$\mu$m)} & \colhead{[3.6]} & \colhead{[4.5]} & \colhead{Class} & \colhead{Region} \\
             & \colhead{(mag)}            & \colhead{(mag)}            & \colhead{(mag)}                    & \colhead{(mag)}           & \colhead{(mag)}             & \colhead{(mag)}           & \colhead{(mag)}           & \colhead{(mag)} & \colhead{(mag)} \\
\colhead{(1)}& \colhead{(2)}              & \colhead{(3)}              & \colhead{(4)}                      & \colhead{(5)}             & \colhead{(6)}               & \colhead{(7)}             & \colhead{(8)}             & \colhead{(9)}   & \colhead{(10)}  & \colhead{(11)}  & \colhead{(12)}
}
\startdata
J052946.87+333816.6      & 18.031$\pm$0.040 & 17.557$\pm$0.044 & 16.572$\pm$0.038 & 14.629$\pm$0.037 & 14.629$\pm$0.036 & 14.629$\pm$0.098 & 14.629$\pm$0.213 & 14.540$\pm$0.042 & 13.581$\pm$0.041 & I  & Eastern \\
SSTGLMCG174.2086-00.0741 & 18.090$\pm$0.016 & 16.480$\pm$0.018 & 15.560$\pm$0.015 &        $-$       &        $-$       &        $-$       &        $-$       & 14.680$\pm$0.057 & 14.259$\pm$0.076 & II & Eastern \\
J053035.04+334408.6      & 16.114$\pm$0.009 & 15.184$\pm$0.006 & 14.825$\pm$0.008 & 14.314$\pm$0.030 & 14.314$\pm$0.044 &      14.314      &      14.314      & 14.234$\pm$0.043 & 13.990$\pm$0.043 & II & Eastern \\
J053050.85+332735.1      & 17.645$\pm$0.014 & 16.167$\pm$0.013 & 15.373$\pm$0.013 & 14.674$\pm$0.034 & 14.674$\pm$0.043 & 14.674$\pm$0.233 &      14.674      & 14.368$\pm$0.053 & 13.796$\pm$0.035 & II & Eastern \\
J053047.75+332740.1      & 14.858$\pm$0.003 & 13.865$\pm$0.002 & 13.361$\pm$0.003 & 12.540$\pm$0.031 & 12.540$\pm$0.024 & 12.540$\pm$0.035 & 12.540$\pm$0.097 & 12.336$\pm$0.037 & 11.570$\pm$0.029 & II & Eastern \\
SSTGLMCG174.2067-00.0695 & 16.178$\pm$0.008 & 15.231$\pm$0.006 & 14.744$\pm$0.007 &        $-$       &        $-$       &        $-$       &        $-$       & 14.282$\pm$0.039 & 13.912$\pm$0.042 & II & Eastern \\
J053047.05+334842.9      & 17.324$\pm$0.010 & 15.845$\pm$0.010 & 14.983$\pm$0.009 & 13.948$\pm$0.030 & 13.948$\pm$0.033 & 13.948$\pm$0.109 & 13.948$\pm$0.032 & 14.171$\pm$0.038 & 13.772$\pm$0.037 & II & Eastern \\
J053047.97+334709.8      & 15.189$\pm$0.004 & 14.227$\pm$0.003 & 13.744$\pm$0.003 & 13.155$\pm$0.024 & 13.155$\pm$0.026 & 13.155$\pm$0.064 & 13.155$\pm$0.032 & 13.089$\pm$0.048 & 12.870$\pm$0.038 & II & Eastern \\
\enddata
\tablecomments{The IR photometric magnitudes of the identified YSO candidates. Column 1 lists the YSO ID from the database of $WISE$ and $Spitzer$. Columns 2--4 list the photometric magnitudes of the 2MASS $J$, $H$, and $K_{\rm s}$ bands. Columns 5--8 list the photometric magnitudes of the $WISE$ $W1$, $W2$, $W3$, and $W4$ bands. Columns 9 and 10 list the photometric magnitudes of the $Spitzer$ IRAC [3.6~$\mu$m] and [4.5~$\mu$m] bands. Column 11 lists the classification of the YSOs. Column 12 lists the region where the YSOs are identified. For the photometric magnitudes without uncertainties, the values presented here are the upper limits.}
\end{deluxetable}
\end{longrotatetable}

\begin{deluxetable}{cccccccccc}
\tabletypesize{\scriptsize}
\tablewidth{0pt}
\rotate
\tablecaption{Properties of Outflows \label{tab:tab5}}
\tablehead{
\colhead{Outflow} & \colhead{$v_{\rm center}$} & \colhead{$\Delta v_{\rm blue}$} & \colhead{$\Delta v_{\rm red}$} & \colhead{$N_{\rm H_2,blue}$}    & \colhead{$N_{\rm H_2,red}$}     & \colhead{$M_{\rm blue}$} & \colhead{$M_{\rm red}$} & \colhead{$P_{\rm out}$}          & \colhead{$E_{\rm out}$} \\
                  & \colhead{(km~s$^{-1}$)}    & \colhead{(km~s$^{-1}$)}         & \colhead{(km~s$^{-1}$)}        & \colhead{(10$^{19}$~cm$^{-2}$)} & \colhead{(10$^{19}$~cm$^{-2}$)} & \colhead{($M_\sun$)}     & \colhead{($M_\sun$)}    & \colhead{($M_\sun$~km~s$^{-1}$)} & \colhead{(10$^{45}$~erg)} \\
\colhead{(1)}     & \colhead{(2)}              & \colhead{(3)}                   & \colhead{(4)}                  & \colhead{(5)}                   & \colhead{(6)}                   & \colhead{(7)}            & \colhead{(8)}           & \colhead{(9)}                    & \colhead{(10)}
}
\startdata
E-F\_OF  &  -3.43 &  [-11.59, -6.92] &    [0.39, 3.97] & 4.35 & 2.43 & 0.79 & 1.39 & 16.81 & 1.28 \\
W-F\_OF1 & -15.22 &         $-$      & [-12.07, -5.79] & $-$  & 2.63 & $-$  & 1.35 & 12.74 & 1.19 \\
W-F\_OF2 & -15.05 & [-22.86, -18.38] &      $-$        & 2.52 & $-$  & 0.95 & $-$  &  7.41 & 0.58 \\
W-F\_OF3 & -15.22 & [-25.24, -18.21] & [-11.98, -6.19] & 3.87 & 3.43 & 1.67 & 2.35 & 37.93 & 3.57 \\
\enddata
\tablecomments{Properties of identified outflows. Column 2 is the velocity of the line center. Columns 3 and 4 are the velocity ranges of the blue and red line wings. Columns 5 and 6 are the mean H$_2$ column densities of the blue and red lobes. Columns 7 and 8 are the masses of the blue and red lobes. Column 9 is the total momentum of the outflow. Column 10 is the total energy of the outflow.}
\end{deluxetable}

\end{document}